\documentclass[aps,twocolumn,prb,tightenlines,floatfix,showpacs]{revtex4}

\usepackage[dvips]{graphicx}
\usepackage[english]{babel}
\usepackage{amsmath}
\usepackage{amssymb}
\usepackage{times}
\newcommand{\bs}{\begin{split}}
\newcommand{\es}{\end{split}}
\usepackage[dvips]{graphicx}
\newcommand{\be}{\begin{equation}}
\newcommand{\ee}{\end{equation}}
\newcommand{\ba}{\begin{eqnarray}}
\newcommand{\ea}{\end{eqnarray}}
\newcommand{\ek}{\epsilon_{\mathbf{k}}}
\newcommand{\Ek}{E_{\mathbf{k}}}
\newcommand{\uk}{u_{\mathbf{k}}}
\newcommand{\vk}{v_{\mathbf{k}}}
\newcommand{\xik}{\xi_{\mathbf{k}}}

\newcommand{\mb}[1]{{\mathbf{#1}}}

\newcommand{\phik}{\varphi_{\mathbf{k}}}

\begin{document}

\title{A Two Energy Gap Preformed-Pair Scenario For the Cuprates: Implications for Angle-Resolved Photoemission Spectroscopy}

\author{Chih-Chun Chien$^{1}$, Yan He$^{1}$, Qijin Chen$^{2,1}$
and K. Levin$^{1}$}

\affiliation{$^1$James Franck Institute and Department of Physics,
University of Chicago, Chicago, Illinois 60637, USA}
\affiliation{$^2$Zhejiang Institute of Modern Physics and Department of
Physics, Zhejiang University, Hangzhou, Zhejiang 310027, China}

\date{\today}

\begin{abstract}
  We show how, within a preformed pair scenario for the cuprate pseudogap, the
  nodal and antinodal responses in angle resolved photoemission spectroscopy
  necessarily have very different temperature $T$ dependences. We examine
the behavior and the contrasting $T$ dependences for a range
of temperatures both below and above 
$T_c$.  Our calculations are based on a fully microscopic $T$-matrix
approach for addressing pairing correlations in a regime where the
attraction is stronger than BCS and the coherence length is anomalously
short.  Previously, the distinct nodal and anti-nodal responses have
provided strong support for the ``two-gap scenario" of the cuprates in
which the pseudogap competes with superconductivity. Instead, our theory
supports a picture in which the pseudogap derives from pairing
correlations, identifying the two gap components with non-condensed and
condensed pairs. It leads to reasonably good agreement with a range of
different experiments in the moderately underdoped regime and we
emphasize that here there is no explicit curve fitting. Ours is
a microscopic rather than a phenomenological theory.
 We
briefly address the more heavily underdoped regime in which the behavior
is more complex.
\end{abstract}


\maketitle
\section{Introduction}
\subsection{Background Literature}

An important dichotomy is emerging in descriptions of the mysterious
pseudogap phase of the cuprates which has resulted in different
theoretical scenarios \cite{Sawatzky}.  At the heart of this dispute is
whether the pseudogap observed in the normal state is derived from the
superconductivity itself or whether it results from a competing, but
somewhat elusive, order parameter.  Experiments (i) which directly study
this anomalous normal phase have provided evidence for both points of
view \cite{ANLPRL,LoramPhysicaC,Kaminski02,Borisenko}.  However, there
is an even larger class of recent experiments (ii) which address the
superconducting phase. These are based on angle resolved photoemission
\cite{Shen2006,ShenNature,Kaminski} and Raman scattering
\cite{LeTacon06,Guyard} as well as scanning tunneling microscopy
\cite{Gomes07,Yazdani2,Seamus,Boyer07}.  They quite generally reveal
that there are two distinct temperature dependences associated with the
behavior of the spectral function and related properties, in the nodal
and antinodal regions of momentum space.  The nodal response appears to
reflect superconducting order whereas the anti-nodal response is much
less sensitive to $T_c$.  For this reason, it is speculated, that the
pseudogap may derive from a competing order parameter.  Finally, there
is a third class of experiments (iii) which probe the behavior as the
system evolves from above to just below $T_c$ and establish that the
transition is clearly second order. Here, for example, one sees
a very smooth evolution of the ARPES response in the anti-nodal
direction 
\cite{arpesstanford_review,arpesanl}. 
Many other properties \cite{ourreview,LeeReview}
which depend on the excitation gap show no clear
signature of $T_c$.  This is generally interpreted as evidence in favor
of a precursor-superconductivity origin to the pseudogap.

It is the last two classes of experiments which are the focus of this
paper. Indeed, there is very little in the theoretical literature which
addresses these phenomena.  Rather the emphasis has been on the ground
state or on the normal, pseudogap phase.  Our goal is to show how to
reconcile, in particular, the experiments of class (ii) with a preformed
pair scenario.  Moreover, it is possible that the arguments presented
here can be viewed as ``modular" in the sense of applying to alternate
precursor superconductivity approaches such as the ``phase fluctuation"
approach \cite{Emery} or the RVB scheme \cite{Vanilla}.  We stress that
there appear to be no counterpart studies of the intermediate
temperature broken symmetry state within the more widely espoused phase
fluctuation scheme \cite{Emery}.  Our explanation of the dichotomy is
built around a picture in which the short coherence length cuprates are
somewhere between BCS and Bose-Einstein condensed (BEC) systems.  This
crossover scheme seems to be gaining in support
\cite{Sawatzky,LeggettNature}, and is now widely studied in the cold
Fermi gases \cite{ourreview,ChenStripes,RFReview}.  Our emphasis here is
on moderately underdoped cuprates where at the lowest temperatures the
spectral properties appear to conform to that of a simple $d$-wave
BCS-like state \cite{ShenNature,Kanigel}.  While the behavior appears to
be much more complex in the heavily underdoped regime, nevertheless,
there is a smooth evolution with doping and all the indications for
distinct nodal and anti-nodal responses are present at moderate
underdoping.  Thus, we feel the same qualitative physics regarding the
origin of the pseudogap is appropriate to both moderately and heavily
underdoped cuprates.

We build on a $d$-wave BCS-like ground state where the variational
parameters are determined in conjunction with a self-consistency
condition for the chemical potential, $\mu$.  This self consistent
treatment of $\mu$ (which is close to but different from
 $E_F$) is
necessary \cite{Eagles,Leggett} to accommodate the relatively short
coherence length of the cuprates.  Our contribution in the past
\cite{ourreview,ComparReview} has been to address the associated finite
temperature behavior within a microscopic, diagram-based 
T-matrix theory.  In earlier papers the anomalous behavior of the Nernst
coefficient and of the optical conductivity were also addressed within
this framework \cite{Tan,Iyengar}, along with other experiments
\cite{ourreview}, including \cite{Chen4} the nature of the specific heat
jump and the behavior of the conductance $dI/dV$.  Moreover, a number of
years ago \cite{Chen4} we presented a description of the spectral
function with special emphasis on how superconducting coherence would be
evident in the presence of a normal state pseudogap.  A central point of
the present paper is to show that these calculations (which predate the
actual experiments \cite{ShenNature,Kanigel} by five years or more),
yield very good semi-quantitative agreement with a wide range of more
recent ARPES experiments without invoking any fitting parameters or
phenomenology.

At the onset, we present the simple physical picture of the different
ARPES spectral gap responses as a function of $\bf{k}$. We note that the
nodal regions are associated with extended gapless states or Fermi arcs
\cite{Kanigel} which are now rather reasonably well understood
\cite{Normanarcs} within a pre-formed pair scenario above $T_c$. Their
collapse below $T_c$ has also been addressed within the present
formalism \cite{FermiArcs}.  One can anticipate (as we find) that the
arcs are sensitive to the onset of the order parameter, which we call
$\Delta_{sc}$, in the same way that a strict BCS superconductor, (which
necessarily has a gapless normal state), is acutely sensitive to the
onset of ordering.  By contrast, the anti-nodal points are not as affected
by passing through $T_c$ because they already possess a substantial
pairing gap in the normal phase.  One will also reach this conclusion by
arguing that it is a corollary of a second order transition. If there is
a difference between the nodal and anti-nodal responses above $T_c$ (as
is implicit in the presence of the Fermi arcs), it must persist, as we
find here, for some range 
 of temperatures below $T_c$.  A key point to
implementing this physical picture is the realization that the
excitation gap which we call $\Delta$ is, at all temperatures (except
strictly $T=0$), different from the order parameter $\Delta_{sc}$.  This
distinction trivially holds in the normal, pseudogap phase.

\subsection{Physical Picture of BCS-BEC Crossover Scenario}

\begin{figure}
  \includegraphics[width=1.3in,clip] {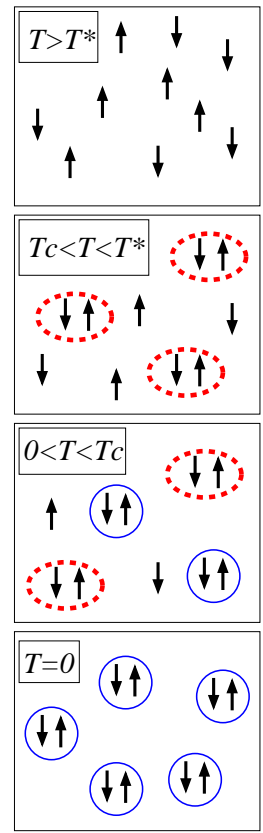} \caption{(Color online)
    Cartoon of the model showing non-condensed pairs in red, open
    ellipses and condensed pairs in blue, closed circles. The number of
    non-condensed pairs scales with the height of the red region in the
    following figure.} \label{Cartoon}
\end{figure}

Before presenting our microscopic scheme it is useful to sketch a simple
physically intuitive approach of the BCS-BEC crossover scenario at
finite temperatures. This approach should be seen to be distinct from
the phase fluctuation scenario. As shown in Figure \ref{Cartoon} the
precursor superconductivity here refers literally to pre-formed pairs,
rather than (as in the phase fluctuation scheme \cite{Emery}) to extended
regions or grains where the order parameter amplitude is well
established while the phase is uncorrelated.
 These pre-formed pairs
arise from a stronger-than-BCS attraction. This strong attraction
breaks the usual
degeneracy between $\Delta$ and $\Delta_{sc}$ or the similar
degeneracy between the pair
formation temperature $T^*$ and condensation temperature $T_c$. Within
this BCS-BEC scenario, the mechanism for pairing need not be specified.
The physics focuses on the anomalously short coherence length of the
cuprates (associated with strong attraction or high $T^*$), whereas in
the phase fluctuation scenario the focus is on the anomalously low
plasma frequency-- leading to soft phase fluctuations, and more
mesoscopic regions of superconductivity.

\begin{figure}
\includegraphics[width=2.5in,clip]
{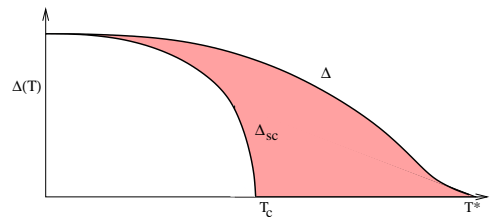}
\caption{(Color online) Contrasting behavior of the excitation gap
  $\Delta(T)$ and superfluid order parameter $\Delta_{sc}(T)$ versus
  temperature.  The number of noncondensed pairs varies as
  $\Delta_{pg}^2 = \Delta^2 - \Delta_{sc}^2$.  }
\label{fig:Delta_Deltasc}
\end{figure}

Figure \ref{Cartoon} shows the schematic behavior as one passes
from above $T^*$ to the fully condensed ground state. The red
(dotted) lines enclose Cooper pairs with net finite momentum,
while the blue (solid) pairs correspond to the components of the
condensate which are at zero center of mass momentum and have 
phase coherence. The third panel with $0<T<T_c$ is the most interesting
from the perspective of the present paper. This is the regime about
which there has been very little theoretical discussion in the
literature and this is the regime where the interesting ``two-gap''
scenario physics is emerging. Here one sees a three-way co-existence: of
the condensate, the fermionic excitations (denoted by a single spin
arrow) and of pair excitations or non-condensed pairs. When there is a
stronger than BCS attractive interaction, preformed pairs above $T_c$,
which are responsible for the pseudogap, do not disappear, but rather
evolve smoothly below $T_c$ into this new form of condensate excitations
arising from non-condensed pairs. This leads to two gap contributions
\cite{Kosztin1} in the superfluid phase representing the finite momentum
pair excitations of the condensate (associated with the component,
$\Delta_{pg}$) and the condensed pairs (associated with the order
parameter, $\Delta_{sc}$).

In this two-gap preformed pair scenario there is a gradual
inter-conversion of non-condensed to condensed pairs as the temperature
decreases. This is shown in Figure \ref{fig:Delta_Deltasc} where the
energy gap parameters are schematically plotted. Above $T_c$ but below
$T^*$ the excitation gap reflects the fact that one has to add energy in
order to create fermionic excitations or break pairs. This excitation
gap $\Delta$ smoothly evolves below $T_c$ as in a second order phase
transition, while precisely at $T_c$ the order parameter $\Delta_{sc}$
opens up. The difference between the (squares) of these two parameters
can be associated with the number of non-condensed pairs. Figure
\ref{fig:Delta_Deltasc} thus shows that the number of non-condensed
pairs is finite below $T_c$ provided the temperature is different from
zero. We will show, using our microscopic scheme that the two gap
components add in quadrature \cite{Kosztin1} to yield the
thermodynamical gap parameter $\Delta(T)$.  Importantly, $\Delta(T)$ is
essentially temperature independent as a consequence of this
inter-conversion from $\Delta_{pg}(T)$ to $\Delta_{sc}(T)$. Just as
there are two gap parameters, there are two temperature scales: $T^*$
marking the gradual onset of the pseudogap, as well as $T_c$ which marks
the appearance of the condensate.

How do we understand the phase diagram of the cuprates within the
BCS-BEC crossover approach? Our interest here is not on the details of
the hole concentration dependence although this has been discussed
elsewhere \cite{ChenStripes,Chen1}.  There is a pronounced competition
between $T^*$ and $T_c$ within the BCS-BEC crossover scenario, as the
attractive interaction $|U|$ increases \cite{NSR}. Indeed, when $T^*$
increases (as for example with underdoping), $T_c$ will ultimately
decrease. This is due to the fact that at large $|U|$, it is
energetically very expensive to unbind a pair of fermions, as is
required in the pair hopping process.  a large effective pair mass is
then responsible for a small $T_c$.  In the $d$-wave case
\cite{ourreview} this pair hopping is even more restricted because of
the extended size of the pair, which leads to pair localization, and
quite possibly the ``singlet glass" phase which has been reported
recently \cite{Seamus}.  Importantly, this concomitant cessation of
$T_c$ occurs while the system is still deep in the fermionic regime
where the chemical potential $\mu$ is positive, suggesting a phase
diagram not so different from that of the cuprates
\cite{Chen2,ChenStripes}.

\section{Overview of Fully Microscopic Theory}

Having discussed the simple physical picture we next review in more
detail the underlying microscopic (T-matrix) theoretical formalism,
which leads to it \cite{ourreview,Chen2,Chen4}.

\subsection{T-matrix Theory}
\label{BCSLeggett}

We begin with a BCS-like ground state: $\Psi_0=\Pi_{\bf k}(\uk+\vk
c_{k,\uparrow}^{\dagger} c_{-k,\downarrow}^{\dagger})|0\rangle$, where
the parameters $\uk$ and $\vk$ are determined variationally in
conjunction with a self-consistent condition for the chemical potential,
$\mu$.  Knowing, as we now do, that at the lowest temperatures the
spectral properties appear to conform to that of simple BCS-like
$d$-wave pairing serves to justify this starting point.  We have
extensively addressed the finite temperature behavior associated with
this fully condensed ground state as well as the spectral properties
\cite{Chen4}.

To address $d$-wave pairing in the cuprates we need to
incorporate specific ${\bf k}$ dependent factors so that
the gap parameters in the self energy acquire the form
$\Delta_{\mb{k},sc}=\Delta_{sc}\phik$
and
$\Delta_{\mb{k},pg}=\Delta_{pg}\phik$,
where we introduce
$\varphi_{\bf k} = \cos (2\phi)$, to reflect the $d$-wave ${\bf k}$
dependence along the Fermi surface.
We adopt a tight binding
model for the band dispersion 
$\ek = 2t(2-\cos k_x -\cos k_y)+2t_z(1- \cos k_z) +4t^{\prime}(1-\cos k_x
\cos k_y) $. 
It should be stressed that all gap parameters have the same
${\bf k}$ dependence. The additional effects of anisotropy 
(beyond those in $\varphi_{\bf k}$) which
appear in the measured spectral gaps,
are not presumed to be present in the initial gap parameters.

We will next briefly summarize the key equations which emerge from our
$T$ matrix scheme \cite{ourreview,ComparReview}.  Throughout this paper,
we adopt a four-vector notation: $Q\equiv (i\Omega_l,\mathbf{q})$,
$K\equiv (i\omega_n, \mathbf{k})$, and $\sum_Q \equiv
T\sum_l\sum_\mathbf{q}$, $\sum_K \equiv T\sum_n\sum_\mathbf{k}$, where
$\omega_n$ and $\Omega_l$ are the odd and even Matsubara frequencies,
respectively. We also take $\hbar=k_B=1$. Within the present approach
there are two contributions to the full $T$-matrix
\begin{equation}
  t = t_{pg} + t_{sc}
\end{equation}
where
\begin{equation}
  t_{sc}(Q)= -\frac{\Delta_{sc}^2}{T} \delta(Q).
\end{equation}
Similarly, we have two terms for the fermion self energy
\begin{eqnarray}
\Sigma(K) &=&
\Sigma_{sc}(K) + \Sigma_{pg} (K)\nonumber\\
 &=& \sum_Q t(Q)G_{0} (Q-K) \varphi_{\bf k -q/2}^2,
\end{eqnarray}
where $G_0$ is the bare Green's function. It follows then that
\begin{equation}
  \Sigma_{sc}(\mb{k},i\omega_n) =
  \frac{\Delta_{\mb{k},sc}^2}{i\omega_n +\ek -\mu} = \frac{\Delta_{\mb{k},sc}^2}{i\omega_n +\xi_\mathbf{k}}.
\end{equation}
Here $\xi_{\bf k} = \epsilon_{\bf k}-\mu$.
Throughout, the label $pg$ corresponds to the ``pseudogap'' and the
corresponding non-condensed pair propagator is given by 
\begin{equation}
t_{pg}(Q)= \frac{U}{1+U \chi(Q)},
\label{eq:14a}
\end{equation}
where the pair susceptibility $\chi(Q)$ has to be properly chosen to
arrive at the BCS-Leggett ground state and $U$ is the attractive pairing
interaction.  We impose the natural BEC condition that below $T_c$ there
is a vanishing chemical potential for the non-condensed pairs
\begin{equation}
\mu_{pair} = 0,
\label{eq:9}
\end{equation}
which means that $t_{pg}(Q)$ diverges at $Q=0$ when $T\le T_c$. Thus, we
approximate \cite{Maly1,Kosztin1} $\Sigma_{pg}(K)$ to yield
\begin{equation}
  \Sigma_{pg} (K)\approx -G_{0} (-K) \Delta_{\mb{k},pg}^2 ,\qquad (T \leq T_c),
\label{eq:sigma3}
\end{equation}
with
\begin{equation}
\Delta_{pg}^2 \equiv -\sum_{Q\neq 0} t_{pg}(Q).
\label{eq:18}
\end{equation}
It follows that we have the usual BCS-like form for the self energy
\begin{equation}
  \Sigma({\bf k}, i\omega_n )  \approx \frac{\Delta_{\bf k}^2}{ i\omega_n
    +\xi_{\bf k}  }, \qquad
  (T \le T_c)
\end{equation}
with $\Delta_{\bf k}=\Delta\varphi_{\bf k}$ and
\begin{eqnarray}
\Delta^2(T) &=& \Delta_{pg}^2 (T) + \Delta_{sc}^2(T).
\label{eq:sum}
\end{eqnarray}
As is consistent with the standard ground state constraints,
$\Delta_{pg}$ vanishes at $T \equiv 0 $, where all pairs are condensed.

Using this self energy, one determines $G$ and thereby can evaluate
$t_{pg}$.  Then the condition that the non-condensed pairs have a
gapless excitation spectrum ($\mu_{pair} =0$) becomes the usual BCS gap
equation, except that it is the excitation gap $\Delta$ and not the
order parameter $\Delta_{sc}$ which appears here.
We then have from
Eq.~(\ref{eq:9})
\begin{equation}
1  + U  \mathop{\sum_{\bf k}}  \frac{1 - 2 f(E_{\bf k})}{2
E_{\bf k}} \varphi_{\bf k}^2  = 0,
\qquad  T \le T_c\,,
\label{eq:gap_equation}
\end{equation}
where $\Ek = \sqrt{\xik^2 +\Delta_{\mathbf{k}}^2}$ is the quasiparticle
dispersion.

To close the loop, for consistency we take for the pair susceptibility
\begin{equation}
\chi(Q)=\sum_{K}G_{0}(Q-K)G(K) \varphi_{\bf k - q/2}^2.
\label{eq:15}
\end{equation}
Here $G = (G_0^{-1} - \Sigma)^{-1}$ is the full Green's function. 
Similarly, using
\begin{equation}
n = 2 \sum_K G(K)
\label{eq:22}
\end{equation}
one derives
\begin{equation}
n  = \sum _{\bf k} \left[ 1 -\frac{\xik}{\Ek}
+2\frac{\xik}{\Ek}f(\Ek)  \right],
\label{eq:23}
\end{equation}
which is the natural generalization of the BCS number equation.
The final set of equations which must be solved is rather simple and
given by Eqs.~(\ref{eq:18}), (\ref{eq:gap_equation}), and (\ref{eq:23}).
Note that in the normal state (where $\mu_{pair}$ is nonzero),
Eq.~(\ref{eq:sigma3}) is no longer a good approximation, although a
natural extension can be readily written down \cite{heyan2}.

To evaluate $\Delta_{pg}^2$ in Eq.~(\ref{eq:18}) we note that at small
four-vector $Q$, we may expand the inverse of $t_{pg}$ after analytical
continuation. Because we are interested in the moderate and strong
coupling cases, where the contribution of the quadratic term in $\Omega$
term is small, we drop this term and thus find the following expression,
which, after analytical continuation, yields the expansion 
\begin{equation}
t_{pg}(Q) = \frac {1}{Z(\Omega - \Omega^0_q+\mu_{pair}) + i \Gamma^{}_Q},
\label{eq:24}
\end{equation}
where
$\Omega^0_\mathbf{q} =
q^2/(2M^*)$
and where $Z$ is the inverse residue given by
\begin{eqnarray}
  Z&=&\frac{\partial t_{pg}^{-1}}{\partial\Omega}\Big|_{\Omega=0,q=0} \nonumber \\
  &=&\frac{1}{2\Delta^2}\left[n-2\sum_{\mathbf{k}}f(\xik)\right].
\label{eq:25}
\end{eqnarray}
We note that the $q^2$ dispersion in $t_{pg}(Q)$ means that for
a range of low $T$, $\Delta_{pg}^2$ will vary as $T^{3/2}$.
We note that, below
$T_c$ the imaginary contribution in Eq. (\ref{eq:24})
$\Gamma^{}_Q \rightarrow 0$
faster than $q^2$ as $q\rightarrow 0$.
It should be stressed that this approach yields the ground state
equations and that it represents a physically meaningful extension of
this ground state to finite $T$.  We emphasize that the approximation in
Eq.~(\ref{eq:sigma3}) is not central to the physics, but it does greatly
simplify the numerical analysis.

\subsection{Detailed Behavior of the Self Energy}

We have seen that,
after analytical continuation, the self energy is given by
$\Sigma(\mb{k},\omega) = \Sigma_{sc}(\mb{k},\omega) +
\Sigma_{pg}(\mb{k},\omega)$, where
\begin{eqnarray}
\Sigma(\mb{k},\omega) &=&
\frac{\Delta_{\mb{k},sc}^2}{\omega +\xik}
+\Sigma_{pg}(\mb{k},\omega) \label{selfE3} \\
&\approx&
\frac{\Delta_{\mb{k},sc}^2}{\omega +\xik} +
\frac{\Delta_{\mb{k},pg}^2}{\omega +\xik}
\label{selfE2}
\end{eqnarray}
The BCS-Leggett ground state equations \cite{Leggett} follow, provided
one makes the approximation contained in Eq.~(\ref{eq:sigma3}). In
invoking this approximation we are in effect ignoring the difference
between condensed and non-condensed pairs which cannot be strictly
correct.  The simplest correction to $\Sigma_{pg}$ (which should apply
above and below $T_c$) is to write an improved form which most
importantly accommodates the fact that the coherent Cooper pairs of the
condensate are infinitely long lived, whereas the incoherent or
non-condensed pairs have a finite inverse lifetime $\gamma$
\begin{equation}
\Sigma_{pg}(\mb{k},\omega) \approx
\frac{\Delta_{\mb{k},pg}^2}{\omega +\xik + i\gamma}
+\tilde{\Sigma} (\mb{k},\omega) \,.
\label{SigmaPG_Model_Eq}
\end{equation}
Here $\tilde{\Sigma} (\mb{k},\omega)$ represents the lifetime associated
with channels other than the pairing channel and, as is conventional, we
parameterize $\tilde{\Sigma} (\mb{k},\omega) \equiv - i \Sigma_0$. Thus
we have
\begin{equation}
\Sigma({\bf k}, \omega)
=
\left(\frac{\Delta_{\mb{k},pg}^2}{\omega+\xik +i\gamma} -i\Sigma_0 \right)
+
\frac{\Delta_{\mb{k},sc}^2}{\omega +\xik}.
\label{eq:1}
\end{equation}

The above equation contains a well known form for $\Sigma_{pg}$. It also
contains the important addition of $\Sigma_{sc}$.  The model for
$\Sigma_{pg}$ was determined in the present context on the basis of
detailed numerical studies \cite{Malypapers,Chen4} and has been deduced
independently \cite{Norman98} and widely applied.  \cite{Normanarcs} in
the cuprate literature.  Here the broadening $\gamma \ne 0$ and
``incoherent'' background contribution $\Sigma_0$ reflect the fact that
noncondensed pairs do not lead to \textit{true} off-diagonal long-range
order.  While we can think of $\gamma$ as a phenomenological parameter
in the spirit of the literature \cite{Normanarcs,Chubukov2} we stress
that there is a microscopic basis for considering this broadened BCS
form \cite{Maly1,Maly2}.  The precise value of $\gamma$, and its
$T$-dependence are not particularly important for the present purposes,
as long as it is non-zero at finite $T$.
By contrast $\Sigma_{sc}$ is associated with long-lived condensed Cooper
pairs, and is similar to $\Sigma_{pg}$ but without the broadening. 
It is, moreover, often assumed that $-i \Sigma_0 \approx -i \gamma$,
although this assumption is not necessary.

\subsection{Spectral function and Superfluid Density}

The resulting spectral function, based on Eq.~(\ref{SigmaPG_Model_Eq})
and Eq.~(\ref{selfE3}) is given by
\be 
A({\bf k},\omega)=\frac{2\Delta_{pg,\mathbf{k}}^{2}\gamma
(\omega+\xi_{\bf k})^2}{(\omega+\xi_{\bf
 k})^2(\omega^2-E_{\bf k}^{2})^2+\gamma^2(\omega^2-\xi_{\bf
 k}^2-\Delta_{sc,\mathbf{k}}^{2})^2} \,.
 \label{spec}
\ee
%
For convenience, here we do
not show the effects of the $\Sigma_0$ term.  
Above $T_c$, Eq.~(\ref{spec}) is used with $\Delta_{sc} = 0$.
It can be seen that at all ${\bf k}$ and below $T_c$, this spectral
function contains a zero at $\omega =-\xi_{\bf k}$, whereas it has no
zero above $T_c$.  This means that a clear signature of phase coherence
is present when one passes from above to below $T_c$, as long as $\gamma
\neq 0$ distinguishes the non-condensed from the condensed pairs.

These dramatic effects of the condensate in the spectral function are
also important for addressing the specific heat jump at $T_c$ which must
be present as a thermodynamic indication of the phase transition. The
onset of a condensate below $T_c$ (with no lifetime broadening, $\gamma
=0$ ) in contrast to the lifetime broadened contribution from the
pseudogap is associated with clear signatures in the specific heat
\cite{Chen4} as the system develops superconducting coherence.

Physically, one can anticipate that the non-condensed pairs represent an
additional mechanism for destroying the condensate.  It is important to
stress that as a consequence this approach is different from a Fermi
liquid based superconductor which has often been presumed in the
theoretical literature \cite{LeTacon06}.  Because the normal state is,
by consensus, a non-Fermi liquid, and because there is a smooth
evolution from above to below $T_c$, it should not appear surprising
that the superconducting phase is also non-Fermi liquid-based.
Important to the analysis of the superfluid density is the imposition of
gauge invariance through a Ward identity. In this way one finds
\cite{Chen2,JS} that the pseudogap contributions via $\Sigma_{pg}$ to
the superfluid density precisely cancel, in contrast to those from
$\Delta_{sc}$.

After this cancellation, the superfluid
density is found to be  of the simple form\cite{Chen2}
\begin{equation}
\left(\frac{n_s(T)}{m}\right) = \left( 1 - \frac{\Delta_{pg}^2(T)}
{\Delta^2(T)}
\right)
\left( \frac{n_s(T, \Delta(T)) }{m} \right) ^{BCS}.
\label{eq:ns}
\end{equation}
Here, importantly, the quantity $(n_s(T, \Delta(T))/m)^{BCS}$ corresponds to
the conventional BCS form for the $d$-wave superfluid density, albeit with an
unusual, essentially $T$-independent gap $\Delta(T)$ in the underdoped regime.
In summary, one sees that $n_s$ is
additionally depressed by \textit{bosonic fluctuations} which insure that
$n_s$ vanishes at $T_c$, not $T^*$.

\subsection{Abbreviated Model}

To make the present formalism more widely accessible we construct a
simplified or abbreviated model in which $T^*$ and $T_c$ are effectively
fit to the cuprate phase diagram and the various gap parameters
$\Delta_{pg}$ and $\Delta_{sc}$ which appear in the spectral function
are then readily deduced.  For the purposes of the present paper we do
not focus on this short cut scheme, but it serves to make the results
here easily reproducible by others.

We have seen that in the temperature regime below or only slightly above
$T_c$, the thermodynamical energy gap $\Delta(T)$ and its component
$\Delta_{pg}(T)$ satisfies $\Delta^2(T) = \Delta_{pg}^2(T) +
\Delta_{sc}^2(T)$ where we define $\Ek^{2d} \equiv \sqrt{ (\xik^{2d}) ^2
  + \Delta _{\bf k}^2}$ and presume that $\Delta(T) \equiv
\Delta_{mf}(T)$ 
satisfies the (two dimensional,
mean field) BCS gap equation
\begin{eqnarray}\label{dgap}
  0 &=& 1  + U  \mathop{\sum_{\bf k}}  \frac{1 - 2 f(\Ek^{2d} )}{2
   \Ek^{2d}} \varphi _{\bf k}^2,~~\mbox{with} \\
  \Delta_{pg}^2 (T) &\approx& \left(T/T_c\right)^{3/2} \Delta^2 (T_c),
  \qquad  T \le T_c\,, \nonumber \\
  &=& \Delta^2(T), \qquad\qquad\qquad\,\,\; T\ge T_c\, .
\label{eq:Dpg}
\end{eqnarray}

\begin{figure}
  \includegraphics[width=3.0in,clip] {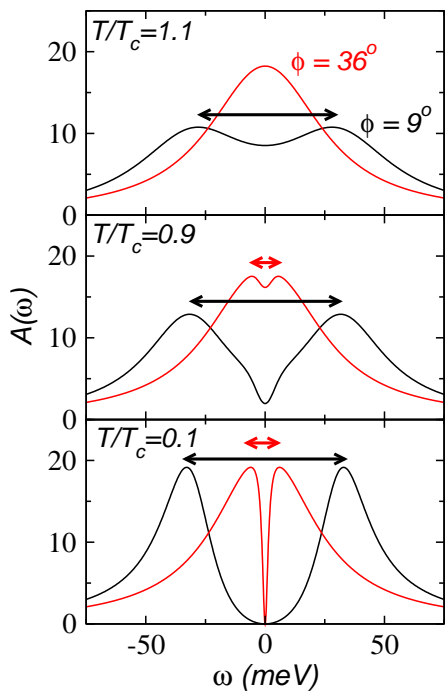}
  \caption{(Color online) Spectral function $A(\phi,\omega)$ at
    $T/T_c=1.1, 0.9, 0.1$ (from top to bottom) for $\phi=9^\circ$
    (black) and $\phi=36^\circ$ (red). Black and red arrows indicate
    size of the spectral gap, which is measured in ARPES. }
\label{fig:Sp}
\end{figure}

\begin{figure}
  \centerline{\includegraphics[width=3.0in,clip] {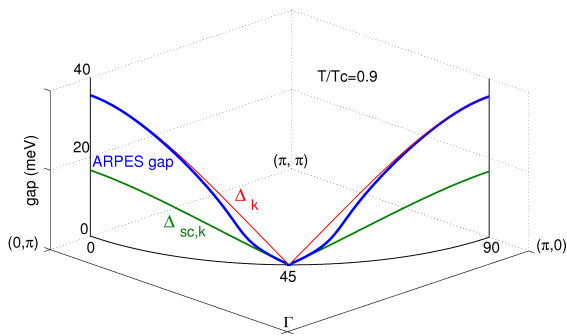}}
\caption {(Color online) ARPES gap (blue thick line),
  $\Delta_{\mathbf{k}}$ (red thin line), and $\Delta_{{\bf k},sc}$
  (green thick line) as a function of $\phi$ at $T/T_c=0.9$.  }
\label{fig:ARPES_3D_gaps}
\end{figure}

\begin{figure}
  \includegraphics[width=3.0in,clip] {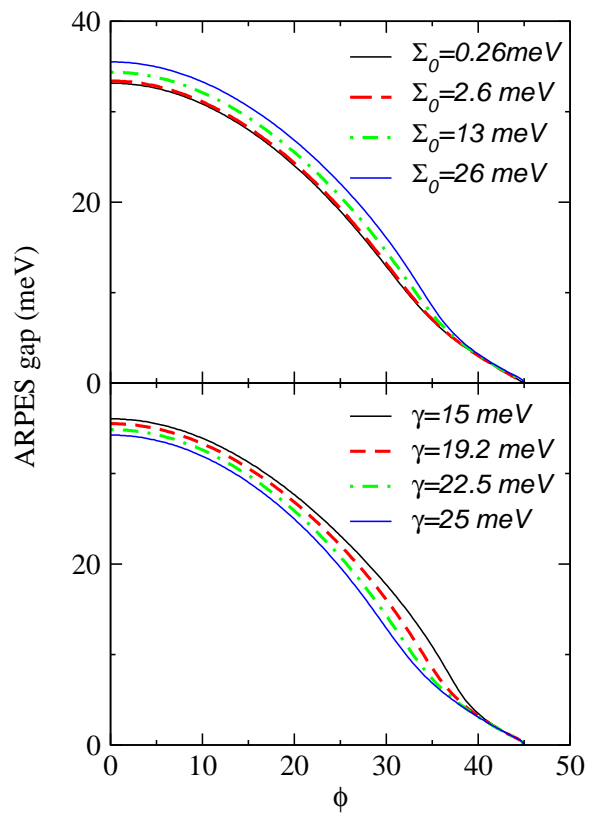} 
  \caption {(Color online) Parameter Insensitivity. This is illustrated
    for $T/T_c = 0.9$. Here we restricted $\gamma$ to produce
    appropriately large arcs in the normal phase. Within this range
    there is virtually no change in the size of the deduced spectral
    gap. We explore two orders of magnitude variation in $\Sigma_0$ and
    again find no change in the spectral gap size.}
\label{Sig0Ga}
\end{figure}

\begin{figure*}[t]
  \centerline{ \includegraphics[width=5.9in,clip] {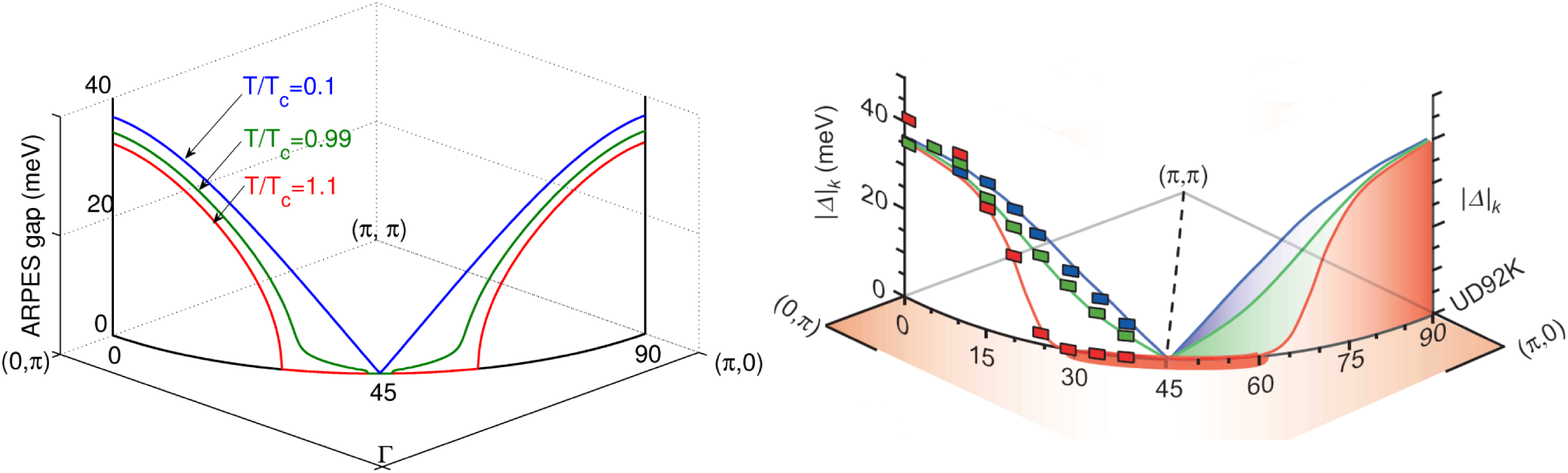}}
\caption{Contrasting nodal and anti-nodal temperature dependences in the
  $d$-wave case.  Figure on the left is the ARPES gap as a function of
  angle $\phi$ at $T/T_c=1.1, 0.99, 0.1$ (labeled on the figure). This
  figure should be compared with the experimental plots on the right
  taken from Figure 4b in Ref.~\onlinecite{ShenNature}}.
\label{fig:ARPES_3D}

\label{Photo3d}
\end{figure*}

Here the superscript $2d$ refers to the fact that we drop the
third dimension in the energy dispersion so that $t_z \rightarrow 0$. 
At each $x$, the parameter $U$ is chosen to yield the measured $T^*$
and, knowing $T_c$, $\Delta(T_c)$ can be determined.  These equations
must be solved in conjunction with a self consistent particle number
equation for $\mu$.  Lying behind this phenomenological approach is the
fact that in a fully consistent theory, \cite{Chen1} $T_c$ is
(logarithmically) 
dependent on the inter-layer hopping $t_z$, and it vanishes when this
parameter is absent where the system is strictly two dimensional. Thus
we can view $t_z$ as a fitting parameter which depends on hole
concentration $x$. In the fully self consistent scheme one recovers the
entire cuprate phase diagram for $T^*(x)$ and $T_c(x)$ by a proper
choice of $U(x)$ and $t_z(x)$.  The short cut scheme then allows one to
calculate without too much effort, the various gap parameters as a
function of temperature and $x$ which appear in the spectral function.

We see that because the total gap $\Delta(T,x)$ satisfies the BCS
equation there is a BCS-relation between $T^*$ and $\Delta(T=0)$.  In
this way Eq.~(\ref{dgap}) implies that the excitation gap $\Delta$
contains the energy scale $T^*$, not $T_c$. Indeed, at intermediate
values of the attractive interaction $|U|$, $\Delta(T)$ is essentially
independent of temperature from the ground state (where $\Delta_{pg}=0$)
to well above $T_c$.  We will not discuss the hole concentration
dependence $x$ in detail in this paper, because it has been treated
elsewhere \cite{Chen2,FermiArcs}.  Finally, we note that within this
BCS-BEC scenario, the mechanism for pairing need not be specified.
Nevertheless, it is clear that the increase of $T^*$ with decreasing $x$
requires that the attractive pairing interaction must become stronger as
the Mott insulator phase is approached.

\section{Numerical Results}
\subsection{General Properties of the Spectral Functions}

We turn now to detailed numerical calculations of the behavior of the
spectral function, $A(\phi,\omega)$ on the Fermi surface (where $\ek-\mu
=0$).  Throughout we will define the spectral (or ARPES) gap as one-half
the peak to peak separation in the spectral function (when it exists).
The
dispersion $\ek$ 
is obtained using our two dimensional tight binding model.  For the most
part we will consider a prototypical hole concentration $x=0.125$, which
is associated with a particular value of $U$ in
Eq.~(\ref{eq:gap_equation}) leading to $T_c/T^* \approx 0.5$. We choose
a bandwidth of $4t=250$ meV and this results in a $T=0$ gap about 34
meV.  Our results are insensitive to the specific parameter set as we
will demonstrate below. The only constraint to be imposed from
experiment is that there must be sizeable Fermi arcs (of order, say,
$10^\circ$ out of $45^\circ$) in the normal phase, for a moderately underdoped
sample.  This means that the parameter $\gamma $ at $T_c$ is not much
less than about one half $\Delta$ at the same temperature.  The
parameter $\Sigma_0$ is found to be relatively unimportant for the
purposes of the plots we present here. It is reasonable to presume that
the lifetime of the non-condensed pairs increases as temperature is
lowered, since their number becomes fewer. For definiteness, following
Ref.~\onlinecite{FermiArcs}, we take $\Sigma_0=26$ meV independent of
$T$ and $\gamma=26$ meV at $95$ K with $\gamma(T) = \gamma(95$ K$)(T/95$
K$)$ above $T_c$ and $\gamma=\gamma(T_c)(T/T_c)^3$ below $T_c$.  To be
more consistent with experimental data, when spectral functions are
presented we convolve the spectral function with a Gaussian instrumental
broadening curve with a standard deviation $\sigma=3$ meV.

Figure~\ref{fig:Sp} illustrates the temperature evolution of the
spectral function for $\phi=9^\circ$ (close to the antinodes) and
$\phi=36^\circ$ (close to the nodes) at $T/T_c=1.1, 0.9, 0.1$ 
from top to bottom.  Above $T_c$ (top panel) the well understood
behavior \cite{Normanarcs,FermiArcs} sets the stage for the normal
phase which underlies the superconducting state in the next two
panels. In this top panel, one sees Fermi arcs, which derive from
the broadening term $\gamma$ in $\Sigma_{pg}$, in the near-nodal
direction, and a pseudogap in the spectral function, associated
with $\Delta_{pg}$ near the anti-nodes. These arcs appear
over that range of $\bf {k}$ values for which $\gamma$ is
larger than the momentum dependent pseudogap. When $T$ is slightly below
$T_c$ (middle panel), a dip in the spectral function at $\phi =
36^\circ$ suddenly appears at $\omega=0$. At this $\phi$ the
underlying normal state is gapless so that the onset of the
additional component of the self energy via $\Sigma_{sc}$ with
long-lived pairs ($\gamma =0$) leads to the opening of a spectral
gap.

By contrast, the presence of this order parameter is not
responsible for the gap near the anti-nodes ($\phi = 9^\circ$),
which, instead, mostly derives from $\Delta_{pg}$.  Here the
positions of the two maxima are relatively unchanged from their
counterparts in the normal phase.  However, $\Delta_{sc}$ does
introduce a sharpening of the spectral function, associated with
the deepening of the dip at $\omega=0$. This can be seen
analytically from Eq.~(\ref{spec}) by noting that $\Sigma_{sc}$
suppresses $A(\omega)$ near $\omega=0$. When $T\ll T_c$ (lower
panel), pairing fluctuations are small so that $\Delta(T) \approx
\Delta_{sc}(T)$ and one returns to a conventional BCS-like
spectral function with well established gaps at all angles except
at the precise nodes.

It is useful to look at the behavior of the ARPES gap over the entire
range of $\phi$, as studied experimentally \cite{ShenNature}.  To
emphasize that the spectral gap does not precisely correspond to the
self energy gap components, in Fig.~\ref{fig:ARPES_3D_gaps} we plot the
spectral function gap along with $\Delta$, and $\Delta_{sc}$ as a
function of angle at $T/T_c=0.9$.  The figure illustrates that, near the
anti-nodes, the spectral gap reflects the magnitude of $\Delta$. Near
the nodes, however, the spectral gap is more directly associated with
$\Delta_{sc}$, in the sense that this gap appears only in the ordered
phase.  The second of these observations is in line with previous
experimental findings \cite{Shen2006,Kaminski}. However, it has
generally been assumed that at the anti-nodes the behavior is governed by
the so-called ``pseudogap". We stress that our interpretation is not at
odds with this literature. Rather we refer to the full gap at the
anti-nodes as $\Delta(T)$ which is roughly a constant in temperature.
This contains two contributions, one from $\Delta_{pg}(T)$ and one from
the order parameter $\Delta_{sc}(T)$. While near $T_c$ the former
dominates, near $T \approx 0$, the latter is the more important. Thus
the gap at the antinodes reflects superconducting order as well, at
least in these moderately underdoped cuprates.

Figure \ref{Sig0Ga} shows that the spectral gap shown by the blue lines
in the previous figure for $T/T_c = 0.9$ is only very slightly modified
when the parameters $\Sigma_0$ (in the top panel) and $\gamma$ (in the
bottom panel) are altered. While the height of the peaks in the spectral
function plots will be affected, the important derived quantities such
as the spectral gap plotted in the figure are not changed when
$\Sigma_0$ is varied by two orders of magnitude. Moreover, if $\gamma$
is reasonably constrained to yield a sizeable Fermi arc above $T_c$,
then the behavior of the spectral gap below $T_c$ 
does not depend on the detailed
values for $\gamma$.

\subsection{Comparison Between Theory and Experiment}

\begin{figure}
\includegraphics[width=2.8in,clip]
{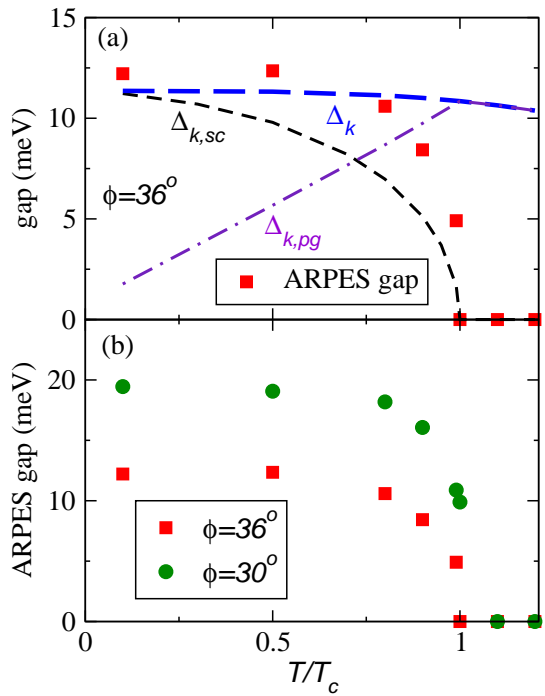}
\caption{(Color online) (a) The ARPES gap (red squares), $\Delta_{\bf
    k}$ (thick blue dashed line), $\Delta_{{\bf k},sc}$ (black dashed
  line), and $\Delta_{{\bf k},pg}$ (orange dot-dash line) as a function
  of $T/T_c$ for $\phi=36^\circ$.  (b) The ARPES gap as a function of
  $T/T_c$ for $\phi=36^\circ$ (red squares) and $\phi=30^\circ$ (green
  circles).  This panel should be compared with Figure 2d of
  Ref.~\onlinecite{ShenNature}.}
\label{fig:gap_T}
\end{figure}

Recently there has been an emphasis on experiments which contrast the
behavior around the gap nodes with that around the gap maxima (or
anti-nodes). The right panel of Figure \ref{Photo3d} indicates the size
of the ARPES or spectral gap as deduced from one-half of the peak to
peak separation in the spectral function. These data \cite{ShenNature}
address a moderately underdoped sample. The three different curves
correspond to three different temperatures with the legend the same as
that in the left panel (representing the results of the present theory.)
Importantly, one sees a pronounced temperature dependence in the
behavior of the ARPES spectral gap for the nodal region (near
45$^\circ$), as compared with the anti-nodal region (near 0 and
90$^\circ$), where there is virtually no $T$ dependence.

Theory (on the left) and experiment (on the right) are in reasonable
agreement and one can readily understand the contrasting temperature
response associated with the different $\bf{k}$ points on the Fermi
surface.  To see this, note that the nodal regions reflect extended
gapless states or Fermi arcs \cite{Kanigel} above $T_c$. It is natural
to expect that they are sensitive to the onset of $\Delta_{sc}$, in the
same way that a strict BCS superconductor, (which necessarily has a
gapless normal state), is acutely sensitive to the presence of order.
By contrast, the anti-nodal points are not as affected by passing
through $T_c$ because they already possess a substantial pairing gap in
the normal phase.

The dramatic variation in the temperature dependence of the spectral gap
as one moves along the Fermi surface has given rise to the so-called
``two gap scenario" \cite{Sawatzky}.  In (perhaps) overly simplistic
terms the one gap and two gap scenarios are differentiated by the
presumption that in the former the pseudogap represents a precursor to
superconductivity, while in the latter the mysterious cuprate pseudogap
is viewed as arising from a competing order parameter. The two gap
scenario is viewed as a consequence of a number of different experiments
\cite{Sawatzky,Seamus} all of which have been interpreted to suggest
that the antinodal region is associated with this alternative (hidden)
order parameter pseudogap and the nodal region is dominated by
superconductivity.  By contrast the viewpoint expressed here (based on
BCS-BEC crossover theory) leads naturally to a different $T$ dependence
for the nodal and anti-nodal region, but at the same time it belongs to
the class of theories which argue that the pseudogap is intimately
connected with the superconductivity.

We turn in Fig.~\ref{fig:gap_T} to very important temperature dependent
studies \cite{ShenNature} which suggest that the nodal gap may directly
reflect the order parameter.  Figure \ref{fig:gap_T}(a) plots the
various gap parameters, $\Delta(T), \Delta_{pg}(T)$ and $\Delta_{sc}(T)$
in the self energy as compared with the spectral gap measured near the
node at $\phi = 36^\circ$ (indicated by squares) as a function of
temperature. It can be seen that this spectral gap, while it is distinct
from the order parameter $\Delta_{sc}(T)$ (except at the lowest
temperatures), vanishes rather close to $T_c$. The figure shows that the
gap parameter $\Delta(T)$ is relatively constant through $T_c$, so that
the decrease in $\Delta_{pg}(T)$ with decreasing $T$ is compensated by
the increase in $\Delta_{sc}(T)$ through the inter-conversion of
non-condensed and condensed pairs. To compare directly with experiment,
in Fig.~\ref{fig:gap_T}(b) we plot the spectral gap for two different
angles, $\phi$, as a function of $T$, in a fashion which looks rather
similar to Fig.~2(d) of Ref.~\onlinecite{ShenNature}.  For $\phi =
30^\circ$, which is somewhat further from the nodes there is a small
spectral gap (pseudogap) above $T_c$.  Because of the $\varphi _{\bf k}$
factor, closer to the anti-nodes the overall magnitude of the ARPES gap
is larger than at $\phi = 36^\circ$.

\begin{figure}
  \includegraphics[width=2.8in,clip] {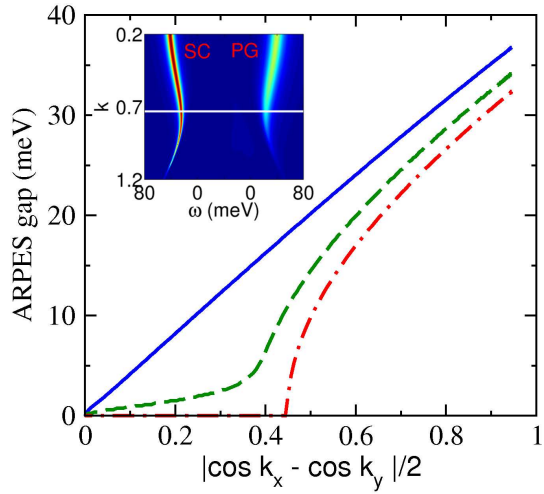}
  \caption { The ARPES gap as a function of $ |\cos(k_x) -\cos(k_y)|/2$
    for $T/T_c=0.1$ (blue solid line), $0.99$ (green dashed line), and
    $1.1$ (red dot-dash line). This should be compared with Figure 3b of
    Ref.~\onlinecite{ShenNature}.  Inset is a contour plot of the
    occupied spectral weight at $\phi = 22.5^o$, showing peak sharpening
    below $T_c$. We follow a similar sweep as that in
    Ref.~\onlinecite{ANLPRL} and the white line indicates the
    intersection with the Fermi surface. Here the intensity
    corresponding to below (left panel) and above (right) $T_c$ is
    largest(smallest) in the red(blue) and we have taken smaller
    $\gamma$ for illustrative purposes. }
\label{fig:contour}
\end{figure}

In Fig.~\ref{fig:contour} we address the important issues which have
been raised in Refs.~\onlinecite{ANLPRL}, \onlinecite{ShenNature}, and
\onlinecite{Kanigel}.  These papers make the case that the pseudogap is
a consequence of the superconductivity.  The figure in the main body is
a plot of the spectral gap for a few different temperatures from above
to below $T_c$ as a function of the simplest $d$-wave form for $\varphi
_{\bf k}$.  This figure compares favorably with Fig.~3(b) in
Ref.~\onlinecite{ShenNature}.  The central point illustrated here is
that at the lowest temperatures one reverts, in effect, to a simple
one-gap scenario. That is, the BCS-like ground state wavefunction
obtains with $\Delta = \Delta_{sc}$.

In the inset of Fig. \ref{fig:contour} we present a contour plot of the
occupied spectral weight corresponding to the product of the spectral
function and Fermi function. In this way one can infer the dispersion
relationship associated with the normal phase and see to what extent it
is related to that below $T_c$.  The left panel is below $T_c$ and the
right panel above $T_c$. This contour plot, albeit represented
differently, compares rather favorably with Fig.~4 in
Ref.~\onlinecite{ANLPRL}.  The similarity of the two panels would not be
expected if the pseudogap were related to another order parameter.

Together Figure~\ref{fig:contour} and related experiments
\cite{ShenNature,Kanigel,ANLPRL} provide evidence that the pseudogap has
to be viewed as ultimately associated with the superconductivity.  The
normal state excitations appear to have a (broadened) BCS-like
dispersion.  The nodal and anti-nodal behavior appear to be intimately
connected in the ground state.

\section{Phenomenological Model for Heavily Underdoped System}

There is a growing body of work on more heavily underdoped cuprates
\cite{Shen2006,Kaminski,Seamus} from which one can infer that the simple
$d$-wave, BCS-like ground state may not be appropriate nearer to the
insulating phase. Here, if one looks at the experimental analogue of
Figure \ref{fig:contour}, the lowest temperature behavior still exhibits
a deviation from the simple $\cos k_x - \cos k_y$ form. Indeed kinks are
often seen \cite{Kaminski} somewhat like that shown in Figure
\ref{fig:contour}, but for the case of very low temperatures. The kinks
are associated with the fact that the ARPES gap curves in the nodal
region seem to reflect the superconducting order while, as before the
antinodal behavior reflects what is referred to as the pseudogap. As a
result it has been argued that \cite{Kaminski} ``the very different
properties of these two gaps lead us to conclude that there is no direct
relationship between the pseudogap and the superconducting gap".

Because there appears to be a rather continuous \cite{ShenNature}
evolution from moderate to heavy underdoping, we, instead speculate that
the physics of the pseudogap in the two regimes must be rather similar
and that the non-simple $d$-wave ARPES gap behavior at the lowest
temperatures in heavily underdoped cuprates is a natural extension of
the higher $T < T_c$ behavior seen at moderate underdoping.  At these
higher $T < T_c$ there are two gap components $\Delta_{pg} \neq 0$ and
$\Delta_{sc} \neq 0$. Thus, a reasonable precursor- superconductivity-
based phenomenological model for this extreme underdoped regime is to
presume that $\Delta_{pg}$ persists into the ground state, perhaps
because of a contamination from the near-by insulating phase. We view
this insulating state as introducing a finite value for the zero
temperature pseudogap.  This is consistent with the way \cite{Chen1} the
insulating phase appears in our calculations where the $pg$ gap
component persists to the lowest temperatures while $\Delta_{sc}$ is
strictly zero beyond a critical value for the attractive interaction, or
equivalently a critical value for $T^*$.

We emphasize that all previous discussions and figures have been
microscopically based and derived, but in this section we proceed purely
phenomenologically. The goal of this discussion is to arrive at a model
for the extreme underdoped case which is smoothly connected to the
physical picture we have thus far exploited for more moderately doped
cuprates. We need to incorporate a clear deviation from the $d$-wave
ground state, kinks or other breaks in the ARPES gap function which
distinguish different gap shapes around the nodal and anti-nodal
regimes, and clear evidence for incoherence even below $T_c$, but only
near the antinodes.  The model we present grew out of a discussion with
A.  Yazdani and his collaborators \cite{Yazdaniprivate} who have
observed a similar gap shape in their STM experiments.
 
\begin{figure}
\includegraphics[width=2.9in,clip]
{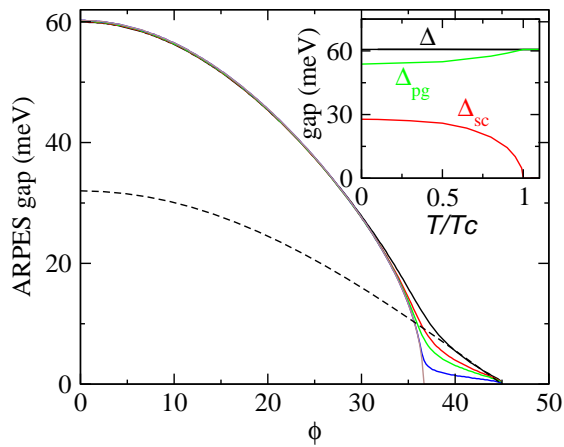}
\caption {(Color online) Behavior of the spectral gap as a function of
  angle $\phi$ for a phenomenological model representing a heavily
  underdoped system.  The inset plots the gap functions which should be
  contrasted with that shown in Figure 7a. The dashed line is the
  extrapolation of the simple $\cos(2\phi)$ behavior found near the gap
  nodes. Solid curves from top to bottom correspond to $T/Tc=0.1, 0.8,
  0.9, 0.99, 1.1$.} \label{12model}
\end{figure}

To describe this class of models we assume that all gap functions
(but not the spectral gaps themselves) have the form $\Delta_{pg,{\bf
k}}=\Delta_{pg}\varphi_{\bf k}$ and $\Delta_{sc,{\bf
k}}=\Delta_{sc}\varphi_{\bf k}$. At a given temperature, the
pseudogap now has two contributions: one from the usual preformed
pairs, which will ultimately go into the condensate at
sufficiently low $T$ and another from the admixture of insulating
state which we view as a ``zero temperature pseudogap". In this
way there is a weak temperature dependence in $\Delta_{pg}$
associated with the pair conversion process and concomitantly
$\Delta_{sc}$ is also $T$ dependent. A typical parameter set is
shown in the inset of Figure \ref{12model}. This plot is to be
contrasted with the behavior shown in Figure \ref{fig:gap_T}a.

For definiteness we presume that the total excitation gap is given by
the mean field gap $\Delta_{mf}(T)$ defined in Eq.~(\ref{dgap}), so that
the superconducting order parameter contribution is $\Delta_{sc}(T)
=\sqrt{\Delta_{mf}^{2}(T)-\Delta_{pg}^{2}(T)}$. The pseudogap
contribution is written as
$\Delta_{pg}(T)=\sqrt{\Delta_{pg0}^{2}(T)+\Delta_{pg1}^{2}(T)}$ with the
zero temperature pseudogap given by
$\Delta_{pg0}(T)=\alpha\Delta_{mf}(T)$ and
$\Delta_{pg1}(T)=(T/Tc)^{3/2}\sqrt{\Delta_{mf}^{2}(T)-\Delta_{pg0}^{2}(T)}$
for $T\le T_c$ and $\Delta_{pg}(T)=\Delta_{mf}(T)$ for $T>T_c$.  Here
$\Delta_{mf}(T)$ is the gap obtained from a mean-field calculation of
$d$-wave BCS theory, as derived from Eq.~(\ref{dgap}).  In our
microscopic calculations one would have $\alpha=0$ which appears
consistent with moderately underdoped systems. However, for heavily
underdoped cuprates We choose $\alpha$ such that the sc and pg
contributions at T=0 are in the ratio of 1:2, as a typical example. We
take $\gamma(T)=\Sigma_0(T)=0.5\Delta_{pg}(T)$ for $T\le T_c$ and
$\gamma(T)=\Sigma_0(T)=(T/T_c)\gamma(T_c)$ for $T>T_c$.

Figure \ref{12model} shows the behavior of the spectral gap for the
heavily underdoped model and for a range of temperatures $T/Tc=0.1, 0.8,
0.9, 0.99, 1.1$. The dashed line is an extrapolation of the simple
$d$-wave fitted form found near the gap nodes and associated with the
order parameter $\Delta_{sc}$ at the lowest temperature. While there is
a simple $d$-wave fitted form also at the anti-nodes the effective gap
here is the much larger parameter $\Delta$, which consists mostly of a
pseudogap contribution, for this heavily underdoped system.

\begin{figure}
  \includegraphics[width=2.9in,clip] {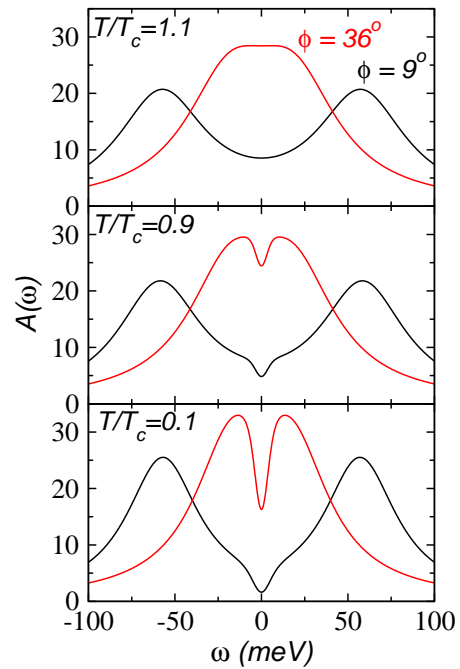}
\caption {(Color online) Spectral functions with convolution for
  phenomenological model of a heavily underdoped system. This model
  should show that at the lowest $T$ the behavior around the antinodes
  is not much more coherent than that in the normal state. This figure
  should be contrasted with Figure~\ref{fig:Sp}.} \label{Sp12}
\end{figure}

Figure \ref{Sp12} shows a plot of the actual spectral functions at two
angles $\phi = 36^o$ in red and $\phi = 9^o$ in black at three different
temperatures from above to just below $T_c$ to finally at $T/T_c =
0.1$. The behavior in this heavily underdoped system can be contrasted
with that shown in Figure~\ref{fig:Sp} for moderate doping. The nodal
curves show the Fermi arc behavior above $T_c$, followed by the opening
of a gap (which reflects superconductivity) below $T_c$ and the ultimate
establishment of well defined coherence with decreased $T$ as evident by
the narrow, well defined peaks. By contrast the anti-nodal regime
(unlike its counterpart in Figure~\ref{fig:Sp}) does not indicate the
presence of coherent quasi-particles. Rather, even at the lowest
temperatures the peaks are broad, and very little changed from those
above $T_c$.

There are features of this model which do not capture all the phenomena
observed experimentally. The ``kink" effects seem to be strictly
associated with the Fermi arcs of the normal state and not particularly
close to the magnetic zone boundary \cite{Shen2006}, since the arc size
is rather small in this underdoped regime.  Moreover we have presumed a
strictly $d$-wave gap shape which constrains the behavior of the
spectral gap near the antinodes.  Nevertheless, this is a reasonable
model for further study, since it does preserve some of the key physics
of the experiments.

\section{Conclusions and Comparisons with the Literature}

This paper addresses issues which are at the center of major debates in
high temperature superconductivity.  Do the recent (so-called ``two gap"
) experiments which report a difference associated with the nodal and
anti-nodal response in ARPES \cite{ShenNature,Shen2006,Kaminski} or in
Raman \cite{LeTacon06,Guyard} or scanning tunneling microscopy
\cite{Seamus,Boyer07} rule out the possibility that the pseudogap derives
from the superconductivity itself? We argue that, despite strong claims
in the literature, pseudogap formation owing to preformed pairs is, in
fact, consistent with these experiments. 
We stress that our approach for the moderately underdoped cuprates is
\textit{not phenomenological}. It was in place well \cite{Chen4}
before these experiments were undertaken.

We have emphasized that our explanation for the physics is relatively
simple and is based on a stronger-than-BCS attractive interaction,
associated with short coherence length Cooper pairs. The formation of
isolated pairs 
(in contrast to extended regions of fixed pairing amplitude) takes place
at $T^*$, while condensation appears at $T_c$.  What is crucial is that
pseudogap effects which are associated with these pre-formed pairs do
not disappear immediately below $T_c$. Rather they persist as
non-condensed pair excitations of the condensate. This is not a Fermi
liquid based form of superconductivity, because there are bosonic
degrees of freedom associated with the fermion pairs.  Nor should this
be thought of as a ``one gap" picture. There are two components to the
pairing gap, one from the non-condensed pairs and another from the
condensate.

A central equation is Eq.~(\ref{eq:1}) which shows that both components
are important in the self energy and therefore in the spectral function.
The contribution from the preformed pairs $\Sigma_{pg}$ is crucial for
forming the Fermi arcs above $T_c$. These appear in the nodal regions
where $\gamma$ is relatively larger than the momentum dependent gap. The
contribution from the condensate $\Sigma_{sc}$ is crucial just below
$T_c$ because it opens up a true gap in the Fermi arc region. This is
reminiscent of a conventional BCS superconductor which necessarily has a
gapless normal state and is, thus, extremely sensitive to the presence
of coherent order. This is, in contrast to the anti-nodal regimes where
the large pseudogap above $T_c$ is very little affected by the addition
of the superconducting order, except through peak sharpening or
coherence effects.

In the context of Eq.~(\ref{eq:1}) it is generally believed
\cite{Normanarcs} that there is only one component to the self energy
($\Sigma_{pg}$) and that the onset of coherence coincides with a
dramatic decrease in $\gamma$ below $T_c$.  We strongly disagree with
this assumption.  Rather there are two contributions to the self energy
below $T_c$ and only one above.  Thus, one should not argue that
$\gamma$ precisely vanishes at $T_c$ but rather there is a continuous
conversion from non-condensed to condensed pairs as $T$ is lowered
within the superfluid phase. The non-condensed pairs below $T_c$ have
finite lifetime while the condensed pairs do not.

In this paper we also discussed the fact that there is support from
another class of experiments that the pseudogap and the superconducting
gap are intimately connected \cite{ShenNature,Kanigel,ANLPRL}. The
lowest temperature spectral properties \cite{ShenNature,Kanigel} of, at
least, moderately underdoped samples seem to fit a simple $d$-wave
angular dependence and recent normal state data \cite{ANLPRL} provide
evidence for a dispersion deduced from the spectral function which is
similar to that in the superfluid phase.

Finally, we addressed heavily underdoped cuprates in a phenomenological
fashion. Here the simple $d$-wave gap shape may not be appropriate
\cite{Kaminski}.  We argued that what is crucial is that there is a
continuous evolution from moderate to extreme underdoping
\cite{ShenNature} so that it is unlikely that the pseudogap has a
different origin in the two regimes. Rather some of the same physics
must be at play. We postulated that there may be a zero temperature
pseudogap present in highly underdoped systems which may derive from
some admixture of the insulating phase.

In summary, this paper has shown how to reconcile a wide class of
experiments in the moderately underdoped cuprates within a pre-formed
pair framework where there are, nevertheless, two components to the
energy gap. This framework \cite{Kosztin1,Chen2} predates the class of
experiments we address here.

This work was supported by Grant Nos. NSF PHY-0555325 and NSF-MRSEC
DMR-0213745. We thank S. Davis, A. Yazdani, Colin Parker and Aakash
Pushp, as well as Wei-Sheng Lee and D. Morr for helpful discussions.

\bibliographystyle{apsrev}

\begin{thebibliography}{44}
\expandafter\ifx\csname natexlab\endcsname\relax\def\natexlab#1{#1}\fi
\expandafter\ifx\csname bibnamefont\endcsname\relax
  \def\bibnamefont#1{#1}\fi
\expandafter\ifx\csname bibfnamefont\endcsname\relax
  \def\bibfnamefont#1{#1}\fi
\expandafter\ifx\csname citenamefont\endcsname\relax
  \def\citenamefont#1{#1}\fi
\expandafter\ifx\csname url\endcsname\relax
  \def\url#1{\texttt{#1}}\fi
\expandafter\ifx\csname urlprefix\endcsname\relax\def\urlprefix{URL }\fi
\providecommand{\bibinfo}[2]{#2}
\providecommand{\eprint}[2][]{\url{#2}}

\bibitem[{\citenamefont{Hufner et~al.}(2008)\citenamefont{Hufner, Hossain,
  Damascelli, and Sawatzky}}]{Sawatzky}
\bibinfo{author}{\bibfnamefont{S.}~\bibnamefont{Hufner}},
  \bibinfo{author}{\bibfnamefont{M.~A.} \bibnamefont{Hossain}},
  \bibinfo{author}{\bibfnamefont{A.}~\bibnamefont{Damascelli}},
  \bibnamefont{and} \bibinfo{author}{\bibfnamefont{G.}~\bibnamefont{Sawatzky}},
  \bibinfo{journal}{Rep. Prog. Phys.} \textbf{\bibinfo{volume}{71}},
  \bibinfo{pages}{062501} (\bibinfo{year}{2008}).

\bibitem[{\citenamefont{Kanigel et~al.}(2008)\citenamefont{Kanigel, Chatterjee,
  Randeria, Norman, Koren, Kadowaki, and Campuzano}}]{ANLPRL}
\bibinfo{author}{\bibfnamefont{A.}~\bibnamefont{Kanigel}},
  \bibinfo{author}{\bibfnamefont{U.}~\bibnamefont{Chatterjee}},
  \bibinfo{author}{\bibfnamefont{M.}~\bibnamefont{Randeria}},
  \bibinfo{author}{\bibfnamefont{M.~R.} \bibnamefont{Norman}},
  \bibinfo{author}{\bibfnamefont{G.}~\bibnamefont{Koren}},
  \bibinfo{author}{\bibfnamefont{K.}~\bibnamefont{Kadowaki}}, \bibnamefont{and}
  \bibinfo{author}{\bibfnamefont{J.~C.} \bibnamefont{Campuzano}},
  \bibinfo{journal}{Phys. Rev. Lett.} \textbf{\bibinfo{volume}{101}},
  \bibinfo{pages}{137002} (\bibinfo{year}{2008}).

\bibitem[{\citenamefont{Tallon and Loram}(2001)}]{LoramPhysicaC}
\bibinfo{author}{\bibfnamefont{J.~L.} \bibnamefont{Tallon}} \bibnamefont{and}
  \bibinfo{author}{\bibfnamefont{J.~W.} \bibnamefont{Loram}},
  \bibinfo{journal}{Physica C} \textbf{\bibinfo{volume}{349}},
  \bibinfo{pages}{53} (\bibinfo{year}{2001}).

\bibitem[{\citenamefont{Kminski et~al.}(2002)\citenamefont{Kminski, Rosenkranz,
  Fretwell, Campuzano, Li, Raffy, Culle, You, Olson, Varma
  et~al.}}]{Kaminski02}
\bibinfo{author}{\bibfnamefont{A.}~\bibnamefont{Kminski}},
  \bibinfo{author}{\bibfnamefont{S.}~\bibnamefont{Rosenkranz}},
  \bibinfo{author}{\bibfnamefont{H.}~\bibnamefont{Fretwell}},
  \bibinfo{author}{\bibfnamefont{J.}~\bibnamefont{Campuzano}},
  \bibinfo{author}{\bibfnamefont{Z.}~\bibnamefont{Li}},
  \bibinfo{author}{\bibfnamefont{H.}~\bibnamefont{Raffy}},
  \bibinfo{author}{\bibfnamefont{W.}~\bibnamefont{Culle}},
  \bibinfo{author}{\bibfnamefont{H.}~\bibnamefont{You}},
  \bibinfo{author}{\bibfnamefont{C.}~\bibnamefont{Olson}},
  \bibinfo{author}{\bibfnamefont{C.}~\bibnamefont{Varma}},
  \bibnamefont{et~al.}, \bibinfo{journal}{Nature}
  \textbf{\bibinfo{volume}{416}}, \bibinfo{pages}{610} (\bibinfo{year}{2002}).

\bibitem[{\citenamefont{Borisenko et~al.}(2004)\citenamefont{Borisenko,
  Kordyuk, Koitzsch, Nenkov, Knupfer, Fink, Grazioli, Turchini, and
  Berger}}]{Borisenko}
\bibinfo{author}{\bibfnamefont{S.}~\bibnamefont{Borisenko}},
  \bibinfo{author}{\bibfnamefont{A.}~\bibnamefont{Kordyuk}},
  \bibinfo{author}{\bibfnamefont{A.}~\bibnamefont{Koitzsch}},
  \bibinfo{author}{\bibfnamefont{K.}~\bibnamefont{Nenkov}},
  \bibinfo{author}{\bibfnamefont{M.}~\bibnamefont{Knupfer}},
  \bibinfo{author}{\bibfnamefont{J.}~\bibnamefont{Fink}},
  \bibinfo{author}{\bibfnamefont{C.}~\bibnamefont{Grazioli}},
  \bibinfo{author}{\bibfnamefont{S.}~\bibnamefont{Turchini}}, \bibnamefont{and}
  \bibinfo{author}{\bibfnamefont{H.}~\bibnamefont{Berger}},
  \bibinfo{journal}{Phys. Rev. Lett.} \textbf{\bibinfo{volume}{92}},
  \bibinfo{pages}{207001} (\bibinfo{year}{2004}).

\bibitem[{\citenamefont{Tanaka et~al.}(2006)\citenamefont{Tanaka, Lee, Lu,
  Fujimori, Fujii, Risdiana, Scalapino, Devereaux, Hussain, and
  Shen}}]{Shen2006}
\bibinfo{author}{\bibfnamefont{K.}~\bibnamefont{Tanaka}},
  \bibinfo{author}{\bibfnamefont{W.~S.} \bibnamefont{Lee}},
  \bibinfo{author}{\bibfnamefont{D.~H.} \bibnamefont{Lu}},
  \bibinfo{author}{\bibfnamefont{A.}~\bibnamefont{Fujimori}},
  \bibinfo{author}{\bibfnamefont{T.}~\bibnamefont{Fujii}},
  \bibinfo{author}{\bibfnamefont{I.}~\bibnamefont{Risdiana},
  \bibfnamefont{Terasaki}}, \bibinfo{author}{\bibfnamefont{D.~J.}
  \bibnamefont{Scalapino}}, \bibinfo{author}{\bibfnamefont{T.~P.}
  \bibnamefont{Devereaux}},
  \bibinfo{author}{\bibfnamefont{Z.}~\bibnamefont{Hussain}}, \bibnamefont{and}
  \bibinfo{author}{\bibfnamefont{Z.~X.} \bibnamefont{Shen}},
  \bibinfo{journal}{Science} \textbf{\bibinfo{volume}{314}},
  \bibinfo{pages}{1910} (\bibinfo{year}{2006}).

\bibitem[{\citenamefont{Lee et~al.}(2007)\citenamefont{Lee, Vishik, Tanaka, Lu,
  Sasagawa, Nagaosa, Devereaux, Hussain, and Shen}}]{ShenNature}
\bibinfo{author}{\bibfnamefont{W.~S.} \bibnamefont{Lee}},
  \bibinfo{author}{\bibfnamefont{I.~M.} \bibnamefont{Vishik}},
  \bibinfo{author}{\bibfnamefont{K.}~\bibnamefont{Tanaka}},
  \bibinfo{author}{\bibfnamefont{D.~H.} \bibnamefont{Lu}},
  \bibinfo{author}{\bibfnamefont{T.}~\bibnamefont{Sasagawa}},
  \bibinfo{author}{\bibfnamefont{N.}~\bibnamefont{Nagaosa}},
  \bibinfo{author}{\bibfnamefont{T.~P.} \bibnamefont{Devereaux}},
  \bibinfo{author}{\bibfnamefont{Z.}~\bibnamefont{Hussain}}, \bibnamefont{and}
  \bibinfo{author}{\bibfnamefont{Z.~X.} \bibnamefont{Shen}},
  \bibinfo{journal}{Nature} \textbf{\bibinfo{volume}{450}}, \bibinfo{pages}{81}
  (\bibinfo{year}{2007}).

\bibitem[{\citenamefont{Kondo et~al.}(2009)\citenamefont{Kondo, Khasanov,
  Takeuchi, Schmalian, and Kaminski}}]{Kaminski}
\bibinfo{author}{\bibfnamefont{T.}~\bibnamefont{Kondo}},
  \bibinfo{author}{\bibfnamefont{R.}~\bibnamefont{Khasanov}},
  \bibinfo{author}{\bibfnamefont{T.}~\bibnamefont{Takeuchi}},
  \bibinfo{author}{\bibfnamefont{J.}~\bibnamefont{Schmalian}},
  \bibnamefont{and} \bibinfo{author}{\bibfnamefont{A.}~\bibnamefont{Kaminski}},
  \bibinfo{journal}{Nature} \textbf{\bibinfo{volume}{457}},
  \bibinfo{pages}{296} (\bibinfo{year}{2009}).

\bibitem[{\citenamefont{Le~Tacon et~al.}(2006)\citenamefont{Le~Tacon, Sacuto,
  Georges, Kotliar, Gallais, Colson, and Forget}}]{LeTacon06}
\bibinfo{author}{\bibfnamefont{M.}~\bibnamefont{Le~Tacon}},
  \bibinfo{author}{\bibfnamefont{A.}~\bibnamefont{Sacuto}},
  \bibinfo{author}{\bibfnamefont{A.}~\bibnamefont{Georges}},
  \bibinfo{author}{\bibfnamefont{G.}~\bibnamefont{Kotliar}},
  \bibinfo{author}{\bibfnamefont{Y.}~\bibnamefont{Gallais}},
  \bibinfo{author}{\bibfnamefont{D.}~\bibnamefont{Colson}}, \bibnamefont{and}
  \bibinfo{author}{\bibfnamefont{A.}~\bibnamefont{Forget}},
  \bibinfo{journal}{Nature Phys.} \textbf{\bibinfo{volume}{2}},
  \bibinfo{pages}{537} (\bibinfo{year}{2006}).

\bibitem[{\citenamefont{Guyard et~al.}(2008)\citenamefont{Guyard, Sacuto,
  Cazayous, Gallais, Le~Tacon, Colson, and Forget}}]{Guyard}
\bibinfo{author}{\bibfnamefont{W.}~\bibnamefont{Guyard}},
  \bibinfo{author}{\bibfnamefont{A.}~\bibnamefont{Sacuto}},
  \bibinfo{author}{\bibfnamefont{M.}~\bibnamefont{Cazayous}},
  \bibinfo{author}{\bibfnamefont{Y.}~\bibnamefont{Gallais}},
  \bibinfo{author}{\bibfnamefont{M.}~\bibnamefont{Le~Tacon}},
  \bibinfo{author}{\bibfnamefont{D.}~\bibnamefont{Colson}}, \bibnamefont{and}
  \bibinfo{author}{\bibfnamefont{A.}~\bibnamefont{Forget}},
  \bibinfo{journal}{Phys. Rev. Lett.} \textbf{\bibinfo{volume}{101}},
  \bibinfo{pages}{097003} (\bibinfo{year}{2008}).

\bibitem[{\citenamefont{Gomes et~al.}(2007)\citenamefont{Gomes, Pasupathy, Ono,
  Ando, and Yazdani}}]{Gomes07}
\bibinfo{author}{\bibfnamefont{K.~K.} \bibnamefont{Gomes}},
  \bibinfo{author}{\bibfnamefont{A.}~\bibnamefont{Pasupathy},
  \bibfnamefont{A~N~Pushp}},
  \bibinfo{author}{\bibfnamefont{S.}~\bibnamefont{Ono}},
  \bibinfo{author}{\bibfnamefont{Y.}~\bibnamefont{Ando}}, \bibnamefont{and}
  \bibinfo{author}{\bibfnamefont{A.}~\bibnamefont{Yazdani}},
  \bibinfo{journal}{Nature} \textbf{\bibinfo{volume}{447}},
  \bibinfo{pages}{569} (\bibinfo{year}{2007}).

\bibitem[{\citenamefont{Pasupathy et~al.}(2008)\citenamefont{Pasupathy, Pushp,
  Gomes, Parker, Wen, Zu, Gu, Ono, Ando, and Yazdani}}]{Yazdani2}
\bibinfo{author}{\bibfnamefont{A.}~\bibnamefont{Pasupathy}},
  \bibinfo{author}{\bibfnamefont{A.}~\bibnamefont{Pushp}},
  \bibinfo{author}{\bibfnamefont{K.}~\bibnamefont{Gomes}},
  \bibinfo{author}{\bibfnamefont{C.}~\bibnamefont{Parker}},
  \bibinfo{author}{\bibfnamefont{J.}~\bibnamefont{Wen}},
  \bibinfo{author}{\bibfnamefont{Z.}~\bibnamefont{Zu}},
  \bibinfo{author}{\bibfnamefont{G.}~\bibnamefont{Gu}},
  \bibinfo{author}{\bibfnamefont{S.}~\bibnamefont{Ono}},
  \bibinfo{author}{\bibfnamefont{Y.}~\bibnamefont{Ando}}, \bibnamefont{and}
  \bibinfo{author}{\bibfnamefont{A.}~\bibnamefont{Yazdani}},
  \bibinfo{journal}{Science} \textbf{\bibinfo{volume}{320}},
  \bibinfo{pages}{196} (\bibinfo{year}{2008}).

\bibitem[{\citenamefont{Kohsaka et~al.}(2008)\citenamefont{Kohsaka, Taylor,
  Wahl, Schmidt, Lee, Fujita, Alldredge, McElroy, Lee, Eisaki et~al.}}]{Seamus}
\bibinfo{author}{\bibfnamefont{Y.}~\bibnamefont{Kohsaka}},
  \bibinfo{author}{\bibfnamefont{C.}~\bibnamefont{Taylor}},
  \bibinfo{author}{\bibfnamefont{P.}~\bibnamefont{Wahl}},
  \bibinfo{author}{\bibfnamefont{A.}~\bibnamefont{Schmidt}},
  \bibinfo{author}{\bibfnamefont{J.}~\bibnamefont{Lee}},
  \bibinfo{author}{\bibfnamefont{K.}~\bibnamefont{Fujita}},
  \bibinfo{author}{\bibfnamefont{J.}~\bibnamefont{Alldredge}},
  \bibinfo{author}{\bibfnamefont{K.}~\bibnamefont{McElroy}},
  \bibinfo{author}{\bibfnamefont{J.}~\bibnamefont{Lee}},
  \bibinfo{author}{\bibfnamefont{H.}~\bibnamefont{Eisaki}},
  \bibnamefont{et~al.}, \bibinfo{journal}{Nature}
  \textbf{\bibinfo{volume}{454}}, \bibinfo{pages}{1072} (\bibinfo{year}{2008}).

\bibitem[{\citenamefont{Boyer et~al.}(2007)\citenamefont{Boyer, Wise,
  Chatterjee, Yi, Kondo, Takeuchi, Ikuta, and Hudson}}]{Boyer07}
\bibinfo{author}{\bibfnamefont{M.~C.} \bibnamefont{Boyer}},
  \bibinfo{author}{\bibfnamefont{W.~D.} \bibnamefont{Wise}},
  \bibinfo{author}{\bibfnamefont{K.}~\bibnamefont{Chatterjee}},
  \bibinfo{author}{\bibfnamefont{M.}~\bibnamefont{Yi}},
  \bibinfo{author}{\bibfnamefont{T.}~\bibnamefont{Kondo}},
  \bibinfo{author}{\bibfnamefont{T.}~\bibnamefont{Takeuchi}},
  \bibinfo{author}{\bibfnamefont{H.}~\bibnamefont{Ikuta}}, \bibnamefont{and}
  \bibinfo{author}{\bibfnamefont{E.~W.} \bibnamefont{Hudson}},
  \bibinfo{journal}{Nature Phys.} \textbf{\bibinfo{volume}{3}},
  \bibinfo{pages}{802} (\bibinfo{year}{2007}).

\bibitem[{\citenamefont{Damascelli et~al.}(2003)\citenamefont{Damascelli,
  Hussain, and Shen}}]{arpesstanford_review}
\bibinfo{author}{\bibfnamefont{R.}~\bibnamefont{Damascelli}},
  \bibinfo{author}{\bibfnamefont{Z.}~\bibnamefont{Hussain}}, \bibnamefont{and}
  \bibinfo{author}{\bibfnamefont{Z.-X.} \bibnamefont{Shen}},
  \bibinfo{journal}{Rev. Mod. Phys.} \textbf{\bibinfo{volume}{75}},
  \bibinfo{pages}{473} (\bibinfo{year}{2003}).

\bibitem[{\citenamefont{Campuzano et~al.}(2004)\citenamefont{Campuzano, Norman,
  and Randeria}}]{arpesanl}
\bibinfo{author}{\bibfnamefont{J.~C.} \bibnamefont{Campuzano}},
  \bibinfo{author}{\bibfnamefont{M.~R.} \bibnamefont{Norman}},
  \bibnamefont{and} \bibinfo{author}{\bibfnamefont{M.}~\bibnamefont{Randeria}},
  \emph{\bibinfo{title}{Physics of Superconductors}}
  (\bibinfo{publisher}{Springer-Verlag}, \bibinfo{address}{Springer, Berlin},
  \bibinfo{year}{2004}), vol.~\bibinfo{volume}{II}, chap.
  \bibinfo{chapter}{Photoemission in the High Tc Superconductors}, pp.
  \bibinfo{pages}{167--273}.

\bibitem[{\citenamefont{Chen et~al.}(2005)\citenamefont{Chen, Stajic, Tan, and
  Levin}}]{ourreview}
\bibinfo{author}{\bibfnamefont{Q.~J.} \bibnamefont{Chen}},
  \bibinfo{author}{\bibfnamefont{J.}~\bibnamefont{Stajic}},
  \bibinfo{author}{\bibfnamefont{S.~N.} \bibnamefont{Tan}}, \bibnamefont{and}
  \bibinfo{author}{\bibfnamefont{K.}~\bibnamefont{Levin}},
  \bibinfo{journal}{Phys. Rep.} \textbf{\bibinfo{volume}{412}},
  \bibinfo{pages}{1} (\bibinfo{year}{2005}).

\bibitem[{\citenamefont{Lee et~al.}(2006)\citenamefont{Lee, Nagaosa, and
  Wen}}]{LeeReview}
\bibinfo{author}{\bibfnamefont{P.~A.} \bibnamefont{Lee}},
  \bibinfo{author}{\bibfnamefont{N.}~\bibnamefont{Nagaosa}}, \bibnamefont{and}
  \bibinfo{author}{\bibfnamefont{X.~G.} \bibnamefont{Wen}},
  \bibinfo{journal}{Rev. Mod. Phys.} \textbf{\bibinfo{volume}{78}},
  \bibinfo{pages}{17} (\bibinfo{year}{2006}).

\bibitem[{\citenamefont{Emery and Kivelson}(1995)}]{Emery}
\bibinfo{author}{\bibfnamefont{V.~J.} \bibnamefont{Emery}} \bibnamefont{and}
  \bibinfo{author}{\bibfnamefont{S.~A.} \bibnamefont{Kivelson}},
  \bibinfo{journal}{Nature} \textbf{\bibinfo{volume}{374}},
  \bibinfo{pages}{434} (\bibinfo{year}{1995}).

\bibitem[{\citenamefont{Anderson et~al.}(2004)\citenamefont{Anderson, Lee,
  Randeria, Rice, Trivedi, and Zhang}}]{Vanilla}
\bibinfo{author}{\bibfnamefont{P.~W.} \bibnamefont{Anderson}},
  \bibinfo{author}{\bibfnamefont{P.~A.} \bibnamefont{Lee}},
  \bibinfo{author}{\bibfnamefont{M.}~\bibnamefont{Randeria}},
  \bibinfo{author}{\bibfnamefont{T.~M.} \bibnamefont{Rice}},
  \bibinfo{author}{\bibfnamefont{N.}~\bibnamefont{Trivedi}}, \bibnamefont{and}
  \bibinfo{author}{\bibfnamefont{F.~C.} \bibnamefont{Zhang}},
  \bibinfo{journal}{J. Phys. - Condens. Matter.} \textbf{\bibinfo{volume}{16}},
  \bibinfo{pages}{R755} (\bibinfo{year}{2004}).

\bibitem[{\citenamefont{Leggett}(2006)}]{LeggettNature}
\bibinfo{author}{\bibfnamefont{A.~J.} \bibnamefont{Leggett}},
  \bibinfo{journal}{Nature Physics} \textbf{\bibinfo{volume}{2}},
  \bibinfo{pages}{134} (\bibinfo{year}{2006}).

\bibitem[{\citenamefont{Chen et~al.}(2007)\citenamefont{Chen, Chien, He, and
  Levin}}]{ChenStripes}
\bibinfo{author}{\bibfnamefont{Q.~J.} \bibnamefont{Chen}},
  \bibinfo{author}{\bibfnamefont{C.-C.} \bibnamefont{Chien}},
  \bibinfo{author}{\bibfnamefont{Y.}~\bibnamefont{He}}, \bibnamefont{and}
  \bibinfo{author}{\bibfnamefont{K.}~\bibnamefont{Levin}}, \bibinfo{journal}{J.
  Supercond. Nov. Magn.} \textbf{\bibinfo{volume}{20}}, \bibinfo{pages}{515}
  (\bibinfo{year}{2007}).

\bibitem[{\citenamefont{Chen et~al.}(2008)\citenamefont{Chen, He, Chien, and
  Levin}}]{RFReview}
\bibinfo{author}{\bibfnamefont{Q.~J.} \bibnamefont{Chen}},
  \bibinfo{author}{\bibfnamefont{Y.}~\bibnamefont{He}},
  \bibinfo{author}{\bibfnamefont{C.-C.} \bibnamefont{Chien}}, \bibnamefont{and}
  \bibinfo{author}{\bibfnamefont{K.}~\bibnamefont{Levin}}
  (\bibinfo{year}{2008}), \bibinfo{note}{eprint, arXiv:0810.1940}.

\bibitem[{\citenamefont{Kanigel et~al.}(2007)\citenamefont{Kanigel, Chatterjee,
  Randeria, Norman, Souma, Shi, Li, Raffy, and Campuzano}}]{Kanigel}
\bibinfo{author}{\bibfnamefont{A.}~\bibnamefont{Kanigel}},
  \bibinfo{author}{\bibfnamefont{U.}~\bibnamefont{Chatterjee}},
  \bibinfo{author}{\bibfnamefont{M.}~\bibnamefont{Randeria}},
  \bibinfo{author}{\bibfnamefont{M.~R.} \bibnamefont{Norman}},
  \bibinfo{author}{\bibfnamefont{S.}~\bibnamefont{Souma}},
  \bibinfo{author}{\bibfnamefont{M.}~\bibnamefont{Shi}},
  \bibinfo{author}{\bibfnamefont{Z.~Z.} \bibnamefont{Li}},
  \bibinfo{author}{\bibfnamefont{H.}~\bibnamefont{Raffy}}, \bibnamefont{and}
  \bibinfo{author}{\bibfnamefont{J.~C.} \bibnamefont{Campuzano}},
  \bibinfo{journal}{Phys. Rev. Lett.} \textbf{\bibinfo{volume}{99}},
  \bibinfo{pages}{157001} (\bibinfo{year}{2007}).

\bibitem[{\citenamefont{Eagles}(1969)}]{Eagles}
\bibinfo{author}{\bibfnamefont{D.~M.} \bibnamefont{Eagles}},
  \bibinfo{journal}{Phys. Rev.} \textbf{\bibinfo{volume}{186}},
  \bibinfo{pages}{456} (\bibinfo{year}{1969}).

\bibitem[{\citenamefont{Leggett}(1980)}]{Leggett}
\bibinfo{author}{\bibfnamefont{A.~J.} \bibnamefont{Leggett}}, in
  \emph{\bibinfo{booktitle}{Modern Trends in the Theory of Condensed Matter}}
  (\bibinfo{publisher}{Springer-Verlag}, \bibinfo{address}{Berlin},
  \bibinfo{year}{1980}), pp. \bibinfo{pages}{13--27}.

\bibitem[{\citenamefont{Levin et~al.}(2008)\citenamefont{Levin, Chen, Chien,
  and He}}]{ComparReview}
\bibinfo{author}{\bibfnamefont{K.}~\bibnamefont{Levin}},
  \bibinfo{author}{\bibfnamefont{Q.~J.} \bibnamefont{Chen}},
  \bibinfo{author}{\bibfnamefont{C.-C.} \bibnamefont{Chien}}, \bibnamefont{and}
  \bibinfo{author}{\bibfnamefont{Y.}~\bibnamefont{He}} (\bibinfo{year}{2008}),
  \bibinfo{note}{eprint, arxiv: 0810.1938}.

\bibitem[{\citenamefont{Tan and Levin}(2004)}]{Tan}
\bibinfo{author}{\bibfnamefont{S.}~\bibnamefont{Tan}} \bibnamefont{and}
  \bibinfo{author}{\bibfnamefont{K.}~\bibnamefont{Levin}},
  \bibinfo{journal}{Phys. Rev. B} \textbf{\bibinfo{volume}{69}},
  \bibinfo{pages}{064510} (\bibinfo{year}{2004}).

\bibitem[{\citenamefont{Iyengar et~al.}(2003)\citenamefont{Iyengar, Stajic,
  Kao, and Levin}}]{Iyengar}
\bibinfo{author}{\bibfnamefont{A.}~\bibnamefont{Iyengar}},
  \bibinfo{author}{\bibfnamefont{J.}~\bibnamefont{Stajic}},
  \bibinfo{author}{\bibfnamefont{Y.~J.} \bibnamefont{Kao}}, \bibnamefont{and}
  \bibinfo{author}{\bibfnamefont{K.}~\bibnamefont{Levin}},
  \bibinfo{journal}{Phys. Rev. Lett.} \textbf{\bibinfo{volume}{90}},
  \bibinfo{pages}{187003} (\bibinfo{year}{2003}).

\bibitem[{\citenamefont{Chen et~al.}(2001)\citenamefont{Chen, Levin, and
  Kosztin}}]{Chen4}
\bibinfo{author}{\bibfnamefont{Q.~J.} \bibnamefont{Chen}},
  \bibinfo{author}{\bibfnamefont{K.}~\bibnamefont{Levin}}, \bibnamefont{and}
  \bibinfo{author}{\bibfnamefont{I.}~\bibnamefont{Kosztin}},
  \bibinfo{journal}{Phys. Rev. B} \textbf{\bibinfo{volume}{63}},
  \bibinfo{pages}{184519} (\bibinfo{year}{2001}).

\bibitem[{\citenamefont{Norman et~al.}(2007)\citenamefont{Norman, Kanigel,
  Randeria, Chatterjee, and Campuzano}}]{Normanarcs}
\bibinfo{author}{\bibfnamefont{M.~R.} \bibnamefont{Norman}},
  \bibinfo{author}{\bibfnamefont{A.}~\bibnamefont{Kanigel}},
  \bibinfo{author}{\bibfnamefont{M.}~\bibnamefont{Randeria}},
  \bibinfo{author}{\bibfnamefont{U.}~\bibnamefont{Chatterjee}},
  \bibnamefont{and} \bibinfo{author}{\bibfnamefont{J.~C.}
  \bibnamefont{Campuzano}}, \bibinfo{journal}{\prb}
  \textbf{\bibinfo{volume}{76}}, \bibinfo{pages}{174501}
  (\bibinfo{year}{2007}).

\bibitem[{\citenamefont{Chen and Levin}(2008)}]{FermiArcs}
\bibinfo{author}{\bibfnamefont{Q.~J.} \bibnamefont{Chen}} \bibnamefont{and}
  \bibinfo{author}{\bibfnamefont{K.}~\bibnamefont{Levin}},
  \bibinfo{journal}{Phys. Rev. B.} \textbf{\bibinfo{volume}{78}},
  \bibinfo{pages}{020513(R)} (\bibinfo{year}{2008}).

\bibitem[{\citenamefont{Kosztin et~al.}(1998)\citenamefont{Kosztin, Chen,
  Jank\'o, and Levin}}]{Kosztin1}
\bibinfo{author}{\bibfnamefont{I.}~\bibnamefont{Kosztin}},
  \bibinfo{author}{\bibfnamefont{Q.~J.} \bibnamefont{Chen}},
  \bibinfo{author}{\bibfnamefont{B.}~\bibnamefont{Jank\'o}}, \bibnamefont{and}
  \bibinfo{author}{\bibfnamefont{K.}~\bibnamefont{Levin}},
  \bibinfo{journal}{Phys. Rev. B} \textbf{\bibinfo{volume}{58}},
  \bibinfo{pages}{R5936} (\bibinfo{year}{1998}).

\bibitem[{\citenamefont{Chen et~al.}(1999)\citenamefont{Chen, Kosztin, Jank\'o,
  and Levin}}]{Chen1}
\bibinfo{author}{\bibfnamefont{Q.~J.} \bibnamefont{Chen}},
  \bibinfo{author}{\bibfnamefont{I.}~\bibnamefont{Kosztin}},
  \bibinfo{author}{\bibfnamefont{B.}~\bibnamefont{Jank\'o}}, \bibnamefont{and}
  \bibinfo{author}{\bibfnamefont{K.}~\bibnamefont{Levin}},
  \bibinfo{journal}{Phys. Rev. B} \textbf{\bibinfo{volume}{59}},
  \bibinfo{pages}{7083} (\bibinfo{year}{1999}).

\bibitem[{\citenamefont{Nozi\`{e}res and Schmitt-Rink}(1985)}]{NSR}
\bibinfo{author}{\bibfnamefont{P.}~\bibnamefont{Nozi\`{e}res}}
  \bibnamefont{and}
  \bibinfo{author}{\bibfnamefont{S.}~\bibnamefont{Schmitt-Rink}},
  \bibinfo{journal}{J. Low Temp. Phys.} \textbf{\bibinfo{volume}{59}},
  \bibinfo{pages}{195} (\bibinfo{year}{1985}).

\bibitem[{\citenamefont{Chen et~al.}(1998)\citenamefont{Chen, Kosztin, Jank\'o,
  and Levin}}]{Chen2}
\bibinfo{author}{\bibfnamefont{Q.~J.} \bibnamefont{Chen}},
  \bibinfo{author}{\bibfnamefont{I.}~\bibnamefont{Kosztin}},
  \bibinfo{author}{\bibfnamefont{B.}~\bibnamefont{Jank\'o}}, \bibnamefont{and}
  \bibinfo{author}{\bibfnamefont{K.}~\bibnamefont{Levin}},
  \bibinfo{journal}{Phys. Rev. Lett.} \textbf{\bibinfo{volume}{81}},
  \bibinfo{pages}{4708} (\bibinfo{year}{1998}).

\bibitem[{\citenamefont{Maly et~al.}(1999{\natexlab{a}})\citenamefont{Maly,
  Jank\'o, and Levin}}]{Maly1}
\bibinfo{author}{\bibfnamefont{J.}~\bibnamefont{Maly}},
  \bibinfo{author}{\bibfnamefont{B.}~\bibnamefont{Jank\'o}}, \bibnamefont{and}
  \bibinfo{author}{\bibfnamefont{K.}~\bibnamefont{Levin}},
  \bibinfo{journal}{Physica C} \textbf{\bibinfo{volume}{321}},
  \bibinfo{pages}{113} (\bibinfo{year}{1999}{\natexlab{a}}).

\bibitem[{\citenamefont{He et~al.}(2007)\citenamefont{He, Chien, Chen, and
  Levin}}]{heyan2}
\bibinfo{author}{\bibfnamefont{Y.}~\bibnamefont{He}},
  \bibinfo{author}{\bibfnamefont{C.-C.} \bibnamefont{Chien}},
  \bibinfo{author}{\bibfnamefont{Q.~J.} \bibnamefont{Chen}}, \bibnamefont{and}
  \bibinfo{author}{\bibfnamefont{K.}~\bibnamefont{Levin}},
  \bibinfo{journal}{Phys. Rev. B} \textbf{\bibinfo{volume}{76}},
  \bibinfo{pages}{224516} (\bibinfo{year}{2007}).

\bibitem[{Mal()}]{Malypapers}
\bibinfo{note}{B. Jank\'o, J. Maly, and K. Levin, Phys. Rev. B \textbf{56},
  R11407, (1997); J. Maly, B. Jank\'o, and K. Levin, Physica C \textbf{321},
  113 (1999).}

\bibitem[{\citenamefont{Norman et~al.}(1998)\citenamefont{Norman, Randeria,
  Ding, and Campuzano}}]{Norman98}
\bibinfo{author}{\bibfnamefont{M.~R.} \bibnamefont{Norman}},
  \bibinfo{author}{\bibfnamefont{M.}~\bibnamefont{Randeria}},
  \bibinfo{author}{\bibfnamefont{H.}~\bibnamefont{Ding}}, \bibnamefont{and}
  \bibinfo{author}{\bibfnamefont{J.~C.} \bibnamefont{Campuzano}},
  \bibinfo{journal}{Phys. Rev. B} \textbf{\bibinfo{volume}{57}},
  \bibinfo{pages}{11093(R)} (\bibinfo{year}{1998}).

\bibitem[{\citenamefont{Chubukov et~al.}(2007)\citenamefont{Chubukov, Norman,
  Millis, and Abrahams}}]{Chubukov2}
\bibinfo{author}{\bibfnamefont{A.~V.} \bibnamefont{Chubukov}},
  \bibinfo{author}{\bibfnamefont{M.~R.} \bibnamefont{Norman}},
  \bibinfo{author}{\bibfnamefont{A.~J.} \bibnamefont{Millis}},
  \bibnamefont{and} \bibinfo{author}{\bibfnamefont{E.}~\bibnamefont{Abrahams}},
  \bibinfo{journal}{\prb} \textbf{\bibinfo{volume}{76}},
  \bibinfo{pages}{180501(R)} (\bibinfo{year}{2007}).

\bibitem[{\citenamefont{Maly et~al.}(1999{\natexlab{b}})\citenamefont{Maly,
  Jank\'o, and Levin}}]{Maly2}
\bibinfo{author}{\bibfnamefont{J.}~\bibnamefont{Maly}},
  \bibinfo{author}{\bibfnamefont{B.}~\bibnamefont{Jank\'o}}, \bibnamefont{and}
  \bibinfo{author}{\bibfnamefont{K.}~\bibnamefont{Levin}},
  \bibinfo{journal}{Phys. Rev. B} \textbf{\bibinfo{volume}{59}},
  \bibinfo{pages}{1354} (\bibinfo{year}{1999}{\natexlab{b}}).

\bibitem[{\citenamefont{Stajic et~al.}(2003)\citenamefont{Stajic, Iyengar,
  Chen, and Levin}}]{JS}
\bibinfo{author}{\bibfnamefont{J.}~\bibnamefont{Stajic}},
  \bibinfo{author}{\bibfnamefont{A.}~\bibnamefont{Iyengar}},
  \bibinfo{author}{\bibfnamefont{Q.~J.} \bibnamefont{Chen}}, \bibnamefont{and}
  \bibinfo{author}{\bibfnamefont{K.}~\bibnamefont{Levin}},
  \bibinfo{journal}{Phys. Rev. B} \textbf{\bibinfo{volume}{68}},
  \bibinfo{pages}{174517} (\bibinfo{year}{2003}).

\bibitem[{Yaz()}]{Yazdaniprivate}
\bibinfo{note}{A. Pushp, C. Parker and A. Yazdani, private communication}.

\end{thebibliography}

\end{document}